%% file: main.tex
\documentclass[manuscript]{acmart}

\usepackage[most]{tcolorbox}

\newtcolorbox{shadedbox}{
  colback=gray!10,
  colframe=black,
  boxrule=0.5mm,
  arc=2mm,
  left=1mm,
  right=1mm,
  top=1mm,
  bottom=1mm
}

\def\Fig{{Fig.}}

\title{Towards AI-Driven Human-Machine Co-Teaming for Adaptive and Agile Cyber Security Operation Centers}

\author{Massimiliano Albanese}
\affiliation{%
  \institution{George Mason University}
  \city{Fairfax}
  \state{VA}
  \country{USA}}
\email{malbanes@gmu.edu}

\author{Xinming Ou}
\affiliation{%
  \institution{University of South Florida}
  \city{Tampa}
  \state{FL}
  \country{USA}}
\email{xou@usf.edu}

\author{Kevin Lybarger}
\affiliation{%
  \institution{George Mason University}
  \city{Fairfax}
  \state{VA}
  \country{USA}}
\email{klybarg@gmu.edu}

\author{Daniel Lende}
\affiliation{%
  \institution{University of South Florida}
  \city{Tampa}
  \state{FL}
  \country{USA}}
\email{dlende@usf.edu}

\author{Dmitry Goldgof}
\affiliation{%
  \institution{University of South Florida}
  \city{Tampa}
  \state{FL}
  \country{USA}}
\email{goldgof@usf.edu}

\begin{document}

\begin{abstract}
Security Operations Centers (SOCs) face growing challenges in managing cybersecurity threats due to an overwhelming volume of alerts, a shortage of skilled analysts, and poorly integrated tools. Human-AI collaboration offers a promising path to augment the capabilities of SOC analysts while reducing their cognitive overload. To this end, we introduce an AI-driven human-machine co-teaming paradigm that leverages large language models (LLMs) to enhance threat intelligence, alert triage, and incident response workflows. We present a vision in which LLM-based AI agents learn from human analysts the tacit knowledge embedded in SOC operations, enabling the AI agents to improve their performance on SOC tasks through this co-teaming. We invite SOCs to collaborate with us to further develop this process and uncover replicable patterns where human-AI co-teaming yields measurable improvements in SOC productivity.
\end{abstract}




\maketitle

\input{introduction}
\input{related_work}
\input{proposed_approach}
\input{framework}
%\input{evaluation}
\input{discussion}

\input{conclusions}
\input{ack}

\bibliographystyle{ACM-Reference-Format}
\bibliography{references}

\end{document}

%% file: introduction.tex
% !TEX root = main.tex

%----------------------------------------------------------------------------------
\section{Introduction} \label{sec:intro}
%----------------------------------------------------------------------------------

Security Operations Centers (SOCs) play a critical role in defending organizations against evolving cyber threats. However, SOC analysts face major challenges, including high alert volumes, repetitive tasks, and insufficient automation. Due to the ever-increasing complexity and interconnectivity of modern networks, the challenges of preventing, detecting, and responding to security incidents have surpassed the current capabilities of SOCs\footnote{In the literature, the terms Cyber Security Operation Center (CSOC) and Security Operation Center (SOC) are often used interchangeably, although the latter is more common}, placing an overwhelming burden on SOC analysts. Real-time threat intelligence is crucial, but the process of collecting, analyzing, and disseminating information can be time-consuming --- as it relies heavily on human analysts to interpret data and make informed decisions --- leaving organizations vulnerable to rapidly evolving threats. Additionally, a shortage of skilled cybersecurity professionals limits the ability of organizations to effectively leverage threat intelligence and respond to threats relevant to them. As these challenges intensify, the demand for more automation in incident response has emerged as a critical focus in cybersecurity research. Leading IT companies are at the forefront of developing tools and orchestration platforms to enhance the efficiency of security incident handling. IBM Resilient, Splunk Phantom, Microsoft Azure Sentinel, Cisco SecureX Orchestration, and Fortinet FortiSOAR are examples of prominent orchestration platforms that support incident response processes. They connect with various security tools and provide a central platform for managing and responding to security incidents. They aim to improve workflows, reduce response times, and increase incident handling efficiency. 

Despite significant advancement in tooling support, SOC analyst burnout~\cite{sundaramurthy2015human} remains a critical issue that was exacerbated by the COVID-19 pandemic~\cite{jones2023work}. Several factors contribute to this challenge. First, the sheer volume of alerts generated by security tools --- most of which are false positives~\cite{alahmadi2022false} --- creates a significant burden for analysts. 
This flood of alerts is often plagued by uncertainty and incomplete data, making it difficult to prioritize and respond effectively.  SOC analysts typically rely on \emph{runbooks} or standard operating procedures (SOP) to address these alerts. However, the repetitive application of these procedures to predominantly false alarms leads to both fatigue and a lack of fulfillment in the role~\cite{sundaramurthy2015human}. Moreover, mainstream security tools often lack the capability to connect important contextual information crucial for comprehending the threat landscape and adapting incident response strategies. Such contextual information may include the specific assets that require protection based on the organization's business needs, the nature of users in the organization's environment, the compliance requirements of the organization, and the political scenario or the economic situation of the organization's country. A human analyst typically considers these contextual factors to decide whether and how to act on an alert. Such reasoning is hard to capture into a typical algorithmic process that can be readily implemented within the traditional framework of building SOC tools. As a result, currently available SOC tools offer limited help to alleviate the pain points in the operations~\cite{sundaramurthy2015human}. Sundaramurthy \emph{et al.} have conducted a long-term anthropological study of SOCs spanning more than a decade and multiple corporate and higher education SOCs~\cite{sundaramurthy2015human,sundaramurthy2016turning,sundaramurthy2014anthropological}. That work showed that, by becoming an anthropologist and embedding within a SOC, a security researcher can extract tacit knowledge within the SOC environment and create tools that {\sf (i)}~readily fit into the analysts' workflow, {\sf (ii)}~precisely address the pain points that result in burnout, and {\sf (iii)}~expand analysts' capability by proactively hunting and responding to threats~\cite{sundaramurthy2016turning}. This approach can be explained by the notion of tacit knowledge --- a concept introduced by Polanyi~\cite{polanyi1966logic,polanyi1966tacit} --- which refers to knowledge held by individuals that has not yet been explicitly articulated. 

\begin{figure}[htbp]
  \centering
  \includegraphics[width=.75\linewidth]{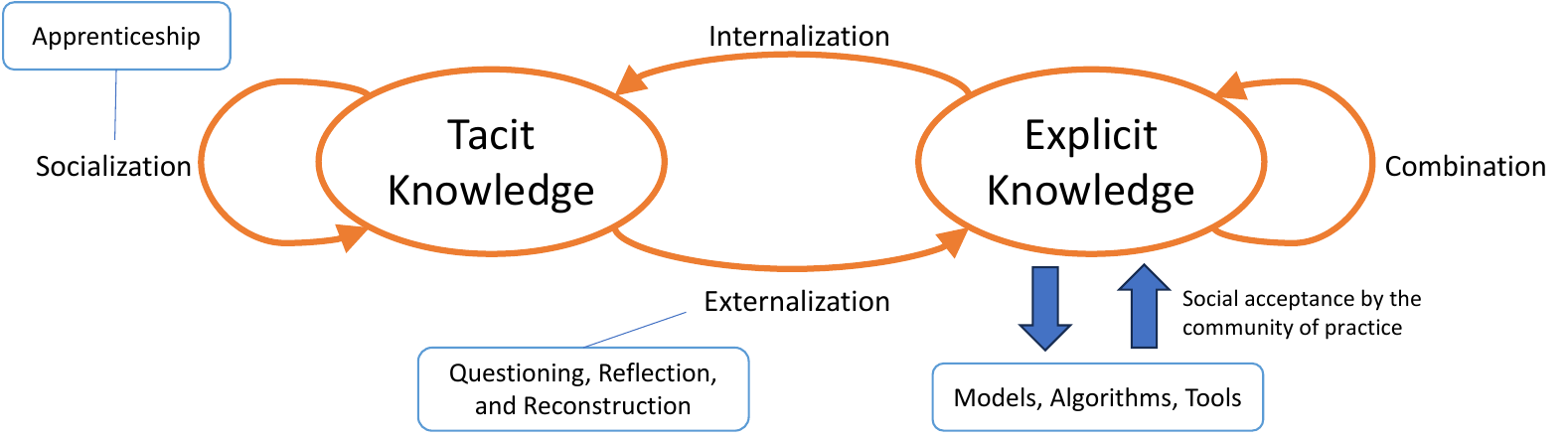}
  \caption{SECI model for anthropology-guided SOC tool building.}
  \label{fig:seci-model}
\end{figure}

The study in~\cite{sundaramurthy2014anthropological} uncovered extensive tacit knowledge among SOC analysts. To access this knowledge, researchers needed to first establish trust with the analysts by embedding themselves in a SOC, performing the same roles as the analysts while observing daily activities through the lens of an anthropologist. This immersive fieldwork, combined with qualitative analysis of fieldnotes, offered deep insights into operational challenges within SOCs that even analysts themselves had not fully recognized. Consequently, the researchers were able to create innovative tools to effectively support the analysts' workflows~\cite{sundaramurthy2015human}. This process can be explained by the SECI knowledge model in \Fig~\ref{fig:seci-model}, first proposed by Nonaka and Takeuchi~\cite{nonaka1995knowledge}. Embedded researchers access tacit knowledge by becoming apprentices of SOC analysts and observing their work as it unfolds (\emph{socialization}). Through questioning, reflection, and reconstruction, the researchers externalize the tacit knowledge into explicit forms such as documentation (\emph{externalization}). The accumulated explicit knowledge is combined (\emph{combination}), internalized (\emph{internalization}), and then applied by the apprentices back to their work. This, in turn, forms new tacit knowledge, and the cycle repeats. Working inside the SOC allows researchers to be part of this knowledge conversion cycle, through which tools readily accepted by the analysts can be developed to codify the explicated knowledge and alleviate the burden on the analysts by automating the most repetitive and mundane part of the investigation task. 

Traditional SOC tools, such as SIEMs, fail to capture the contextual and tacit knowledge necessary for effective decision-making. To address these limitations, we propose an AI-driven human-machine co-teaming approach that enhances SOC workflows through adaptive AI agents. This research builds upon our prior anthropological studies of SOCs, where researchers embedded within security teams uncovered the importance of tacit knowledge in cybersecurity operations. We extend this work by integrating Large Language Models (LLMs) to serve as AI apprentices, assisting analysts in real-time threat investigations.

The remainder of the paper is organized as follows. Section~\ref{sec:related} presents relevant related work, whereas Section~\ref{sec:approach} introduces our vision and overall approach for human-machine co-teaming in SOCs. Then, Section~\ref{sec:framework} presents the proposed framework in detail. Section~\ref{sec:discussion} discusses representative case studies and outlines directions for future research. Finally, Section~\ref{sec:conclusions} concludes the paper, summarizing our contributions.

%% file: related_work.tex
% !TEX root = main.tex

%----------------------------------------------------------------------------------
\section{Related Work} \label{sec:related}
%----------------------------------------------------------------------------------

While still emerging, the application of LLMs in cybersecurity shows promise in tasks like penetration testing~\cite{happe2023penetration,deng2024pentestgpt,shashwat2024pentesting}, incident response, vulnerability management, policy generation, security training, and detecting threats, malware, and intrusions. Research has primarily focused on classification tasks like identifying network intrusions and detecting malicious URLs~\cite{liu2024plm,shenoy2024prompt,chen2024survey,hasanov2024llm}, with some exploration into synthesizing data into unified reports~\cite{siracusano2023cti}, including integrating global threat intelligence with local organizational knowledge~\cite{mitra2024localintel}. However, research on leveraging LLMs to conduct SOC analysis and generate actionable outputs, such as creating or refining incident response plans~\cite{hays2024incident}, remains limited. While there are several cybersecurity-annotated datasets for training machine learning (ML) models on specific tasks~\cite{almhiqani2020insider,kumar2020intrusion,chen2023diversevul,georgescu2020nlpcyber,bayer2024cysecbert}, limited work has focused on instruction tuning to distill expert knowledge and enable diverse cybersecurity-specific tasks~\cite{levi2024cyberpal}. Furthermore, the study of continuous learning strategies in this domain remains largely unexplored. Real-world adoption faces challenges, including privacy concerns with proprietary, cloud-based models like ChatGPT, the high computational demands of LLMs, and the need for cybersecurity-specific models that can capture and distill analysts' knowledge into actionable insights aligned with specific environments and the requirements of individual SOCs~\cite{liu2024plm,shenoy2024prompt,chen2024survey,hasanov2024llm}. Additionally, the probabilistic nature of LLMs introduces variability that contrasts with the deterministic tools SOC analysts typically rely on, raising questions about reliability and trust when identical inputs yield different outputs. This issue is compounded by ``hallucinations,'' where LLMs may provide inaccurate or fabricated information, creating concerns about their dependability for SOC operations~\cite{chen2024survey,hasanov2024llm}. 

We argue that these perceived deficiencies of LLMs are not insurmountable obstacles to their adoption in SOCs. Rather, they arise from attempting to use generative AI models as traditional algorithmic tools. By adopting a human-centric approach and drawing parallels between human SOC analysts and LLMs, we can view an LLM-based AI agent as a collaborator rather than a deterministic tool~\cite{happe2023penetration}. A SOC analyst can use the AI agent to elicit ideas on how to proceed with investigations. The agent can provide supporting evidence for the suggestions, the validity of which human analysts can verify. This human-machine co-teaming is fundamentally different than how an analyst uses traditional SOC tools such as SIEMS. Human analysts do not blindly trust the output of the AI agent but rather must be convinced. The value provided by an AI agent is its ability to churn through vast amounts of data much faster than a human while navigating the nuanced semantics of the data like a human. For investigating cyber incidents, once the relevant data pieces are presented to a human, it is often quite clear what hackers have done and to what effect. However, finding the relevant data pieces across sheer volumes of data may easily overwhelm a human. A machine's upper bound on such bandwidth is much higher than a human's brain, but it must be capable of handling the intricate reasoning to \emph{connect the dots} in the data. How human analysts connect the dots can hardly be specified in a fixed set of rules, and such knowledge is often tacit and hard to explain even by analysts themselves. LLMs present an opportunity to learn such tacit knowledge through data, leading to a human-machine co-teaming where human analysts are gradually liberated from the mundane tasks of processing vast numbers of tickets most of which are false positives~\cite{alahmadi2022false}. 

Geoffrey Hinton highlighted parallels between human cognition and AI systems~\cite{knight2023hinton}. Like LLMs, human analysts are not immune to inconsistencies, errors, and knowledge gaps. However, SOC analysts mitigate these limitations through rigorous training that blends cybersecurity expertise with organization-specific practices. This training includes explicit knowledge such as regulatory compliance requirements, as well as tacit knowledge gained through experience such as recognizing anomalous patterns in network behavior or interpreting subtle contextual cues from threat intelligence reports. LLMs, like human analysts, can improve their reliability through iterative learning and feedback. By integrating mechanisms to capture both explicit and tacit knowledge, LLMs could evolve into valuable SOC collaborators, supporting analysts in a dynamic threat environment.
When working with an LLM-based AI agent, analysts should critically evaluate its suggestions, just as they would evaluate those of a human colleague, requiring evidence-based justification. Viewing an AI agent as a partner rather than a deterministic tool enables SOC teams to utilize its assistance in much the same way they would rely on a human apprentice. Techniques such as prompt engineering, supervised fine-tuning, reinforcement learning from human feedback (RLHF), and retrieval-augmented generation (RAG) can facilitate this adaptation~\cite{yang2024harnessing,fan2024rag}. We envision AI-driven human-machine co-teaming in SOCs, where SOC analysts work alongside AI agents, collaboratively tackling investigation tasks. These agents will enhance analysts' capabilities by internalizing the tacit knowledge required to process nuanced data and help human analysts scale up their work substantially.

In summary, Human-AI collaboration in cybersecurity has been explored in multiple domains, including automated alert triage, threat hunting, and decision support. Existing research on SOC automation highlights the limitations of rule-based systems and the need for adaptive AI-driven approaches. Our work differentiates itself by employing a human-centric co-teaming framework, where LLMs learn from real-world SOC workflows through iterative feedback and adaptation.

% \note{
% \textbf{KL: Brainstorming}
% \begin{itemize}
%     \item \textbf{SOC Operations and Challenges}
%     \begin{itemize}
%         \item Data heterogeneity, alert overload, and tacit knowledge in SOCs
%     \end{itemize}
%     \item \textbf{AI in Cybersecurity}
%     \begin{itemize}
%         \item LLM-assisted threat intelligence, alert triage, and incident response
%     \end{itemize}
%     \item \textbf{Anthropological and Organizational Insights}
%     \begin{itemize}
%         \item The SECI model and human factors influencing SOC performance
%     \end{itemize}
%     \item \textbf{Identified Gaps in Literature}
%     \begin{itemize}
%         \item Limitations in explainability, trust, adaptation, and integration with human workflows
%     \end{itemize}
% \end{itemize}
% }

%% file: proposed_approach.tex
% !TEX root = main.tex

%----------------------------------------------------------------------------------
\section{Vision Overview} \label{sec:approach}
%----------------------------------------------------------------------------------

\begin{figure}[htbp]
  \centering
  \includegraphics[width=\linewidth]{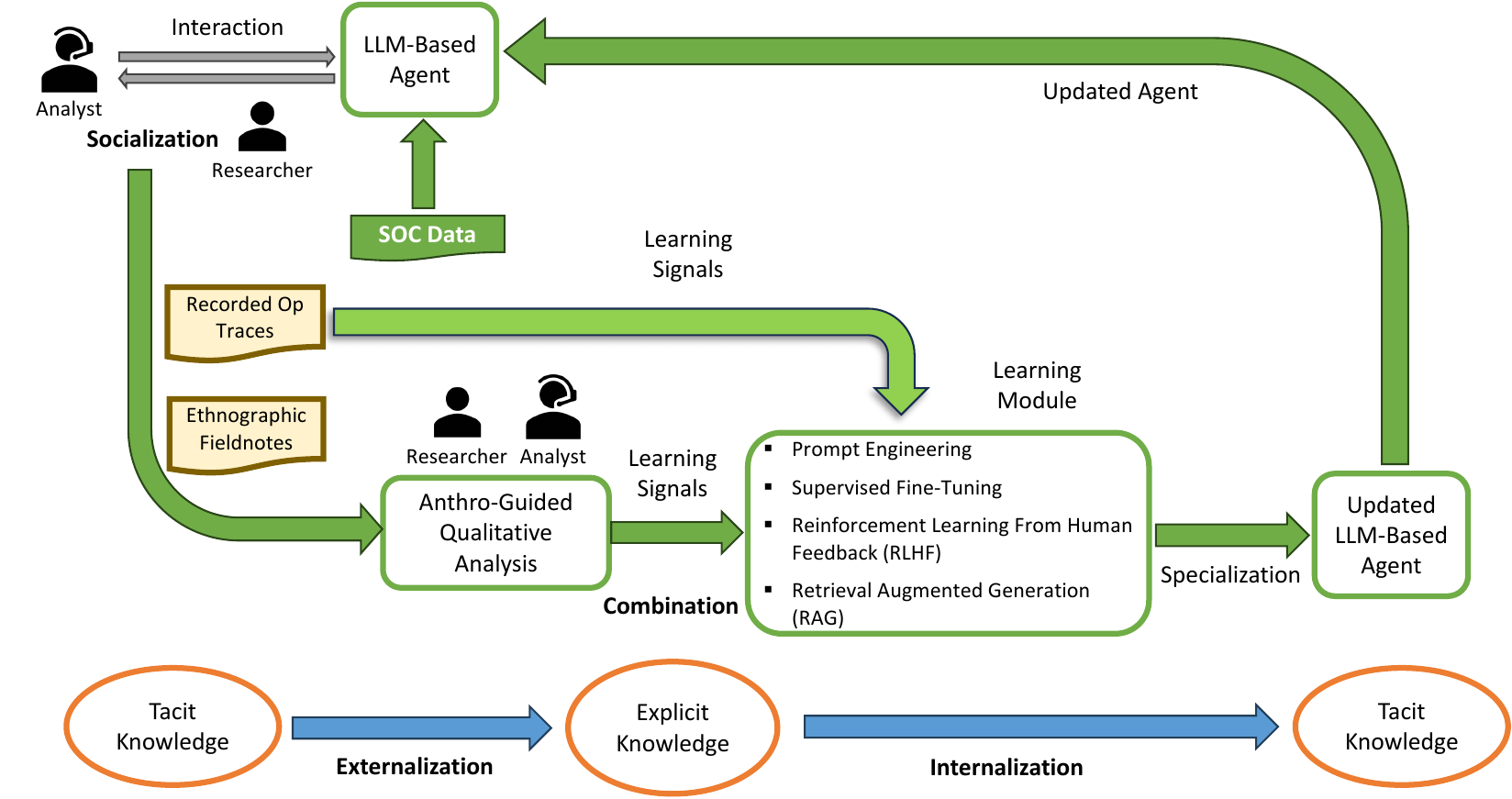}
  \caption{Our vision for human-machine co-teaming for SOC operations.}
  \label{fig:vision}
\end{figure}

Our vision for human-machine co-teaming in SOC operations, partly inspired by our anthropological study of SOCs, is illustrated in \Fig~\ref{fig:vision}. We propose an LLM-based AI agent that acts as an apprentice to human SOC analysts, assisting through interactions with subtasks or investigative ideas. The human analyst will evaluate the agent's outputs. If the results are inaccurate or unhelpful, interactions could provide additional guidance or refute the LLM agent's answer with reasons. A human apprentice would have acquired tacit knowledge through such interactions; for an LLM agent, we need a specific learning process to capture such insights and incorporate the knowledge for the future. This learning process will result in an LLM agent specialized to handle the SOC tasks effectively.

The Learning Module utilizes learning signals derived from analyst-LLM interactions. Op traces consist of raw transcripts of conversations between the analyst and the LLM agent. The ethnographic fieldnotes are recorded by a researcher trained in anthropological methods and embedded in the SOC. This researcher observes interactions between the analyst and the LLM, capturing contextual information in fieldnotes that will be analyzed using qualitative research methods to reveal the tacit knowledge hidden in such interactions. Qualitative analysis can extract classifications, assessments, and higher-order concepts that analysts deem important in their work and, through coding, transform this data into annotations~\cite{lende2023cocreation,fetters2019content}. These codes can describe the category or concept, provide relevant examples, and include inclusion and exclusion criteria (all basic parts of qualitative data analysis)~\cite{thomas2006inductive,bernard2016analyzing}. In the initial stages of the learning process, the embedded researcher can also provide a further source of input about gaps between human preferences and AI outputs, which will provide a triangulation between the SOC, the LLM agent, and the research team. Importantly, this triangulation is based on direct observation rather than secondary assessment, thus increasing the validity of what is working and what is missing in ongoing interactions between analysts and the LLM. Op traces, fieldnotes (including scalar descriptions of what is relevant and non-relevant), and analytic codes from anthropological research can all inform learning signals. These learning signals align the LLM agent's behaviors with the analyst's and the SOC's specific goals. 

Given the sensitivity of SOC data, LLMs used in this research must be housed on premises, necessitating the use of open-source pre-trained foundation models (e.g., Falcon~\cite{almazrouei2023falcon}, LLaMA~\cite{touvron2023llama}, Mistral~\cite{jiang2023mistral}). These models are trained on broad and publicly available corpora, encompassing general web data, books, and technical documents, which may include some cybersecurity content; however, they lack the domain-specific expertise and operational context required for effective SOC operations. Our prior study shows that proficiency in SOC operations requires specialized tacit knowledge gained through experience. Thus, our proposed learning process is based on data generated through the AI agent's experience in handling real SOC tasks. This is a critical step in specializing the agent to assist analysts effectively. The agents will \emph{learn on the job}, just as how human SOC analysts do, so their behavior will gradually align with the specific needs of the SOC. In this online learning paradigm, the agents will require less human input to produce useful results. The cycle can be seen as an instantiation of the SECI model (see \Fig~\ref{fig:seci-model}) in the context of human-LLM co-teaming. It equips the LLM agent with the required tacit knowledge to perform the tasks specific to the SOC, developing a robust human-AI partnership to enhance the effectiveness and efficiency of SOC operations, reduce the cognitive load on analysts, and mitigate burnout.

%----------------------------------------------------------------------------------
\subsection{Learning Strategies} \label{sec:learning_strategies}
%----------------------------------------------------------------------------------

Given the constraints that require LLM agents to operate in on-premises environments, it is essential to explore lightweight learning strategies that avoid the intensive resource demands of training new foundation models. Given the rapid development of LLMs and the emergence of new reasoning capabilities, there are many open questions regarding the tasks that these models can perform without any domain- or task-specific training. In this paradigm, where LLMs approach all language processing tasks as generative tasks, a range of learning strategies becomes relevant. Across tasks, our approach begins with prompt engineering, which provides LLM-specific natural language instructions on how to perform a task, serving as the foundational step in all experimentation. By crafting precise prompts, we can evaluate the baseline capabilities of a given LLM, such as Meta's LLaMA, for specific tasks without additional training. This provides critical insight into whether an LLM's inherent capabilities are sufficient or if further refinement is necessary. For tasks where baseline performance proves inadequate, we can employ supervised fine-tuning~\cite{yang2024harnessing,min2023nlp}, which involves curating a dataset of input-output pairs tailored to task requirements. This allows the model to specialize in specific tasks by training on the curated dataset. Additionally, we can leverage generative feedback~\cite{jin2023alignment}, a cutting-edge approach where users provide natural language critiques of system outputs to guide improvements. This feedback enables users to articulate desired refinements and also provides a mechanism for systematically enhancing the LLM by addressing its current limitations. For more nuanced refinements, particularly in tasks involving subjective evaluation criteria, we can integrate reinforcement learning from human feedback (RLHF)~\cite{yang2024harnessing,min2023nlp}, which aligns LLM outputs with human preferences by using feedback such as rankings or scores in fine-tuning. In this process, the researcher ranks multiple system outputs (e.g., A is better than B) to guide the model towards outputs better aligned with human preferences and expert judgment. RLHF enables iterative improvement by embedding human-like preferences into the model's decision-making processes. Finally, when tasks require integrating external or internal knowledge sources, we can utilize retrieval-augmented generation (RAG)~\cite{fan2024rag}, which integrates external sources to retrieve relevant resources based on user queries and incorporate them into outputs, grounding responses in a predefined knowledge base and ensuring that outputs are informed by up-to-date and domain-specific data. This layered strategy, progressing from prompt engineering to fine-tuning, RLHF, and RAG, provides a flexible, scalable framework to effectively tackle project subtasks and achieve desired outcomes.

%----------------------------------------------------------------------------------
\subsection{Overarching Research Questions}
%----------------------------------------------------------------------------------

Our work is driven by the following two overarching research questions.

\begin{shadedbox} 

\textbf{Research Question 1}.  What processes can be designed to enable LLM agents to effectively learn from experience and improve their performance on SOC tasks? 

\textbf{Research Question 2}. To what extent can these learning processes enhance LLM agents' performance, resulting in productivity gains that substantially outweigh the effort invested in their development? 

\end{shadedbox}

Our preliminary research on applying LLM in software security pen-testing~\cite{shashwat2024pentesting} shows promise for the first research question, so we will leverage this work in our effort. Specifically, after a few iterations of the learning process, we can compare the updated LLM agent's performance against the previous version of the LLM agent on new tasks, utilizing metrics that reflect the agent's helpfulness. One metric could be the number of rounds needed for an analyst to obtain useful information to further the investigation. Our research aims to design specific metrics for an LLM agent based on the specific tasks it is assigned to work on. For the second research question, we can conduct human subject research to obtain feedback from the SOC analysts and use both qualitative and quantitative assessments to measure the return on investment of adopting the LLM agents.

%% file: framework.tex
\section{Framework} \label{sec:framework}
%----------------------------------------------------------------------------------

\begin{figure}[htbp]
  \centering
  \includegraphics[width=\linewidth]{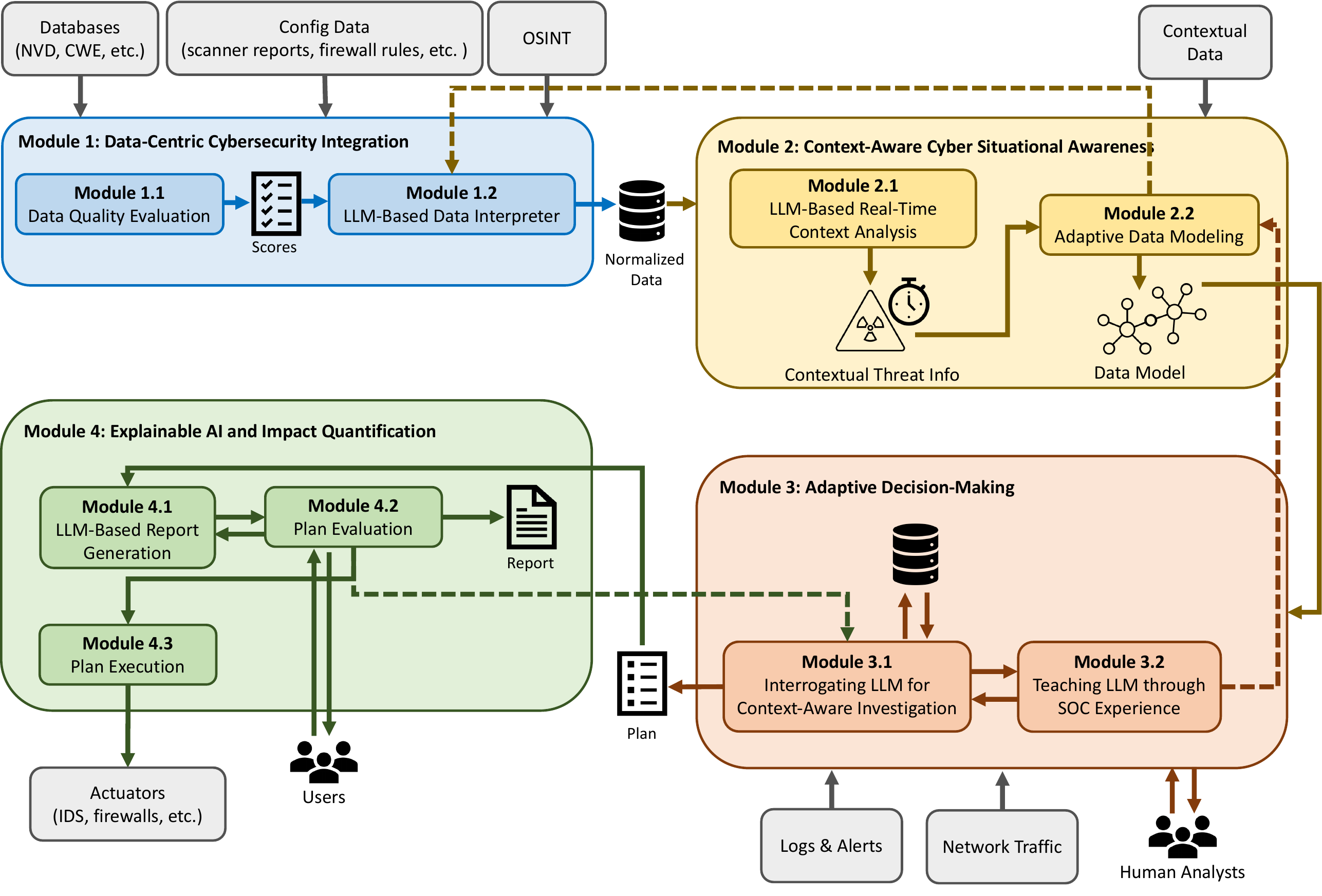}
  \caption{Overview of the proposed framework.}
  \label{fig:framework}
\end{figure}

\Fig~\ref{fig:framework} provides an overview of our framework for human-machine co-teaming, based on the vision laid out in Section~\ref{sec:approach}. The framework comprises four primary modules, each comprising several submodules. 

The framework adopts a data-centric approach that emphasizes cybersecurity integration and
interoperability across diverse tools (Module 1). To encourage
adoption, the framework complements existing tools and capabilities rather
than requiring organizations to overhaul their current systems. By
building on established processes, organizations can
adopt and integrate the framework with minimal disruption. A key
aspect of this approach involves assessing the quality and
trustworthiness of data from various sources (Module 1.1). A key
innovation is the data normalization mechanism, which leverages LLMs
to harmonize data across formats, ensuring tool interoperability and
feeding into the next step: context-aware cyber situational awareness
(Module 2), which includes an internal graphical data model (Module
2.2). This model is grounded in prior work on attack modeling and
configuration security~\cite{albanese2018graphical,albanese2023framework} and provides
input to an adaptive decision-making process (Module 3), capturing
detailed information about security artifacts and their
interrelationships.
Module 3 utilizes an LLM agent and incorporates data from both Module
1 and Module 2 to analyze and interpret logs, alerts, and network
traffic, generating optimal courses of action in response to security
events. The LLM agent dynamically adapts to the evolving threat
landscape by incorporating real-time context analysis of
organizational, economic, and geopolitical data (Module 2.1). This
adaptive data modeling improves the accuracy and timeliness of threat
assessments by contextualizing security events within their broader
environment. Self-awareness mechanisms, based on previous feedback
received by analysts in different but analogous situations, enable AI agents to recognize when human expertise is needed ---
whether to validate findings or contribute to critical decision-making
steps. The final plan of action developed through joint human-machine
collaboration is passed to the Explainable AI and Impact
Quantification module (Module 4). This module utilizes LLMs to
generate a human-readable report detailing the recommended course of
action (Module 4.1) and its impact on the organization's security
posture. The AI-generated explanation is evaluated by various
user groups (Module 4.2), and the feedback loops back into
the adaptive decision-making process to refine future recommendations.
Once validated, the plan is executed by leveraging LLM
capabilities from Module 1 to translate it into instructions for the
organization's security tools (Module 4.3).
The following sections elaborate each module and outline remaining research directions.

%----------------------------------------------------------------------------------
\subsection{Module 1: Data-Centric Cybersecurity Integration} \label{sec:module-1}
%----------------------------------------------------------------------------------

Module 1 adopts a data-centric approach to address the challenges of integrating a vast and growing array of data sources, platforms, and tools to support interoperability and enable large-scale automation.

\begin{shadedbox} \textbf{Research Questions}. How can LLMs be
leveraged to address the challenges of integrating diverse
cybersecurity tools within SOCs, particularly by minimizing
configuration complexities, overcoming the inflexibility of existing
tools, and ensuring interoperability? Additionally, how can these
solutions remain robust against incomplete or uncertain data, as well
as evolving data formats and APIs?
\end{shadedbox}

Modern SOCs handle data in a range of formats used by vulnerability scanners,
log analyzers, intrusion detection systems, and external threat
intelligence sources. The lack of standardization across these tools
presents significant challenges for integration and analysis. This
module explores strategies to minimize complexity, thus
lowering barriers to adoption, and making the proposed solution robust
to incomplete and uncertain data. Given that standardization remains a
primary obstacle to integration, we argue that --- given the complexity
of the problem at hand and the current availability of tools from a
large number of vendors --- pursuing standardization efforts is not the
most effective and efficient way to address the problem, due to
several practical challenges like resistance from vendors, dynamic and
diverse nature of the cybersecurity domain, regulatory and compliance
requirements across industries and regions. To tackle this objective,
the framework includes a metrics-based approach for assessing the
quality and trustworthiness of data gathered from various
sources. An LLM-based mechanism then addresses data
interoperability challenges, focusing on the variability in threat and
vulnerability report formats and the prevalent use of natural language
to describe critical details.

%----------------------------------------------------------------------------------
\subsubsection{Module 1.1: Evaluation of Data Quality and Trustworthiness}
%----------------------------------------------------------------------------------

The primary objective of this module is to establish a robust
mechanism for evaluating the quality, reliability, and trustworthiness
of data ingested from diverse cybersecurity tools and platforms. As
SOCs increasingly depend on inputs such as network logs, vulnerability
scanners, and external threat intelligence, accurately assessing the
reliability and completeness of these data sources is
essential. Initially, data from various sources is normalized
using the mechanisms from Module 1.2. At this stage, the
system operates under a neutral trust assumption, treating all
sources as equally reliable due to the absence of prior knowledge
about their accuracy. Over time, as analysts provide feedback on
recommended actions through human-machine cooperation (Module 3) and
plan evaluation (Module 4.2), this input is used to infer and
adjust the reputation of each data source. It is possible to create a reputation
scoring system based on a centrality metric approach similar to
PageRank~\cite{brin1998search}, where data sources and validated plans
are represented as nodes and feedback is encoded as edges. Unlike
traditional PageRank, where all edges transfer importance, this
approach allows feedback to be positive or negative, reflecting
whether a source contributes accurate or misleading data to a plan,
and uses a bipartite graph with two classes of nodes --- sources and
plans. Scores dynamically adjust based on historical performance
and analyst feedback, enabling the system to prioritize reliable
sources and refine its trust assessments iteratively. This adaptive
scoring system allows continuous improvement as real-world outcomes and
feedback are integrated, enhancing both data quality and the
effectiveness of decision-making within SOC workflows.

%----------------------------------------------------------------------------------
\subsubsection{Module 1.2: LLM-Based Data Interpretation}
%----------------------------------------------------------------------------------

An LLM-based agent interprets and
normalizes data generated by diverse cybersecurity tools and
platforms into a common internal format, translating between
formats for seamless interoperability.

\noindent \textit{Common data format}.  The diverse data formats include structured,
semi-structured, and unstructured data, including natural language
descriptions. A JSON format is used to normalize
cybersecurity threat and vulnerability data, building on the widely
used STIX standard~\cite{barnum2012stix}. This format leverages
key-value pairs, which are both machine-readable and interpretable by
LLMs, enabling the extraction and normalization of diverse data types,
including categorical variables (e.g., severity levels, attack
vectors) and unstructured text (e.g., threat descriptions). The format
integrates data quality and trustworthiness scores from Module 1.1,
to propagate these metrics to the Module 2 graph modeling.

\noindent \textit{Data set curation}.  To develop and evaluate an LLM agent,
the framework curates a parallel corpus of records that span a range of input-output
format combinations. This corpus includes
input-output pairs created using existing format conversion tools to
provide opportunistic data and manually curated samples spanning
format conversions not addressed through available tools to broaden
the format conversions. Vendor-provided data-format specifications are incorporated into the LLM
instructions to ground the format conversion. These specifications
serve as a knowledge source and update path for the LLM agent as
formats change or new formats are released.

\noindent \textit{LLM development}. The LLM development is  treated as a machine
translation task, where the LLM translates between data formats. The
input includes the current and target format, with the LLM trained
to output the data in the desired format. We hypothesize that
vendor-provided format specifications will improve performance,
generalizability, and handling of unseen formats. The framework accommodates multiple representations (raw text, summaries, structured metadata) for incorporating these specifications to determine which most effectively boosts performance.

\noindent \textit{Learning from feedback}.  Within this module, in-context learning
(prompt-based approaches) and supervised fine-tuning are adopted to develop
high-performing LLM agents. Normalized data generated by the agent
support the construction of the graph model in Module 2, which
feeds into Modules 3 and 4. To refine the LLM agent further, the framework incorporates feedback mechanisms into these subsequent modules, including
strategies like RLHF, which uses human feedback to fine-tune models,
aligning outputs with desired outcomes. In this context, RLHF iteratively improves the interpreter by leveraging cybersecurity
analysts’ evaluations of its normalized outputs’ accuracy, relevance,
and completeness.

%----------------------------------------------------------------------------------
\subsection{Module 2: Context-Aware Cyber Situational Awareness} \label{sec:module-2}
%----------------------------------------------------------------------------------

Module 2 provides advanced solutions for comprehensive and
context-aware cyber situational awareness in SOCs and is focused on the following research question.

\begin{shadedbox} \textbf{Research Questions}. How can SOCs achieve
comprehensive, context-aware cyber situational awareness to enable
more informed incident response decisions? How can the context be
expanded beyond organizational boundaries to incorporate cyber threat
intelligence from external sources and additional dimensions such as
economic and geopolitical factors influencing malicious actors?
\end{shadedbox}

Traditional approaches focus on internal assets (\textit{knowledge of
us}) and partial knowledge of adversaries (\textit{knowledge of
them}). The framework extends situational awareness beyond
organizational boundaries to include external factors --- such as
economic conditions, geopolitical tensions, and social dynamics ---
that influence the motivations and behaviors of malicious actors,
ranging from cyber criminals to nation states. 
Diverse threat intelligence sources are aggregated and structured
to create a more holistic view of the cyber threat landscape. By
incorporating this multi-dimensional data, SOCs can adapt their
defense strategies to evolving threats. 
Contextual awareness can also improve both proactive and reactive
cybersecurity measures, such as predicting attack vectors and
prioritizing responses based on real-world implications. This module
includes two sub-modules: Module 2.1 focuses on automatically
analyzing diverse information streams, including news media, social
media, and threat intelligence, to identify SOC-relevant contexts,
assess risk, and identify potential
targets~\cite{trubowitz2021geopolitical}; Module 2.2 aims to integrate
all available information into a comprehensive graphical model.

%----------------------------------------------------------------------------------
\subsubsection{Module 2.1: LLM-based Real-Time Context Analysis}
%----------------------------------------------------------------------------------

This module provides real-time context analysis capabilities
for SOCs by extending Topic Detection and Tracking (TDT) to integrate
dynamic risk assessment. Originally a DARPA-sponsored natural language
processing (NLP) task, TDT focuses on identifying and tracking
emerging events in information streams~\cite{allan1998topic}. The framework builds on TDT to create a near-real-time cybersecurity risk
assessment framework that integrates LLMs and enables continuous
monitoring of diverse data sources, such as news, social media, and
threat intelligence feeds, facilitating the transition from event
detection to actionable situational awareness tailored to an evolving,
SOC-specific threat landscape. In traditional TDT, identifying and
tracking emerging real-world events involves generating structured
event representations using predefined ontologies with attributes and
relations (e.g., trigger/action, location, or target) or clustering
and linking text data to discover and monitor topics, without relying
on explicit
structuring~\cite{xiang2019survey,asgari2021topic,cantini2022topic,akhgari2022sem}. We
can leverage the language understanding and reasoning capabilities of
LLMs through Chain of Thought prompting strategies~\cite{wei2022chain}
to focus on ontology attributes relevant to risk assessment. These
attributes guide the LLM to consider the contextual information most
relevant to SOCs, without requiring strict, predefined templates. For
instance, when tracking a geopolitical event, the LLM can be directed
to assess statements of hostility (trigger), geographic indicators
(location), and the potential threat recipient (target). This
attribute-driven focus helps the LLM prioritize relevant details,
filter out extraneous information, and identify emergent risks.

The risk posed by an identified event to a specific SOC is framed
as a classification problem. Inputs include risk assessment criteria,
the event annotation ontology, a representation of the target, and a
summary of the event. Outputs are categorical risk labels (e.g.,
no threat, low, medium, or high) indicating the perceived threat
level. These labels inform Module 2.2's adaptive model for
dynamic threat assessment. An effective classifier requires a
comprehensive SOC representation that accounts for geopolitical,
sociopolitical, and technical contexts and generalizable criteria for
distinguishing risk levels. It must distinguish between irrelevant
content, relevant but non-threatening content, negative commentary
lacking genuine risk, and content signaling legitimate threats.  The
Sony Pictures Entertainment cyber-attack 2 highlights the potential of
real-time context analysis for enhancing SOC operations. The framework would process media coverage of the film and North Korea’s
hostile response, identifying these stories as both relevant and
threatening. TDT captures North Korea’s public condemnation of
the film, particularly in state-run news outlets labeling it
anti-propaganda~\cite{jong2015interview}. By leveraging TDT ontology
attributes, the LLM could focus on attributes related to the
trigger/action (condemnation), sentiment (threatening or hostile),
actor (North Korea), and target (Sony Pictures) in assessing relevancy
and threat potential. Combining TDT annotation ontology and LLM
language understanding yields a framework capable of identifying
threats while filtering out irrelevant content, such as negative movie
reviews or exaggerated social media posts. By prioritizing actionable
insights, the system supports timely and informed responses.  This
LLM-based approach to TDT produces actionable event summaries and
assigns risk labels for the Adaptive Data Modeling in Module 2.2. This
enables SOC analysts to monitor emerging risks in real time and adapt
defensive strategies in response to geopolitical tensions or other
contextual factors, enhancing situational awareness and enabling
timely responses. Validation is accomplished through retrospective analysis of prior incidents to uncover patterns and
indicators relevant to threat prediction. Using data repositories like
the Center for Strategic and International Studies’ Significant Cyber
Incidents list~\cite{csis2025cyberincidents}, representations for the applicable SOC are generated, utilizing historical media
from tools like the Way Back
Machine~\cite{internetarchive2025wayback}. Analyzing this data
refines the predictive capabilities of real-time systems, helping SOCs
adapt to evolving global contexts. The real-time context analysis
framework also includes feedback loops from subsequent tasks to
refine the LLM-based TDT modeling, further enhancing the accuracy of
context-aware incident response.

%----------------------------------------------------------------------------------
\subsubsection{Module 2.2: Adaptive Data Modeling}
%----------------------------------------------------------------------------------

The goal of Module 2.2 is to develop a dynamic and adaptive data modeling capability that can inform Module 3, supporting real-time risk assessment and effective prioritization of incident response efforts in SOCs. This submodule builds upon the data integration capabilities of Module 1 and the context-aware cyber situational awareness of Module 2.1, using a multi-graph model to capture system components, vulnerabilities, configurations, and external risk factors (e.g., geopolitical conditions). The novelty of this task lies in designing a scalable, continuously updated data model that adapts to the ever-evolving threat landscape and informs automated decision-making processes. The framework adopts a vulnerability-centric approach to data modeling, to enable prioritization of threat responses by focusing on exploitable weaknesses, improving real-time situational awareness and adaptive defenses. This capability builds upon our previous work in developing similar graphical data models~\cite{albanese2018graphical,soroush2020sciborg} and metrics to accurately assess the impact of vulnerability exploits in complex systems~\cite{albanese2023framework,iganibo2023attack} and reason about optimal metrics-driven mitigation strategies. Building upon this body of work, we design a multi-graph model, with subgraphs representing different classes of artifacts, including vulnerabilities, system components, configuration parameters, and external risk factors. These risk factors, identified via Module 2.1, include but are not be limited to geopolitical and economic conditions, with nodes representing geopolitical events or economic sanctions, for example, that can influence the likelihood of attacks like nation-state-driven cyber espionage; compliance and regulatory requirements with nodes representing compliance mandates (e.g., GDPR, HIPAA) mapped to configuration or system component nodes to reflect how failure to meet regulatory changes impact system security; and other factors identified through various threat intelligence feeds (e.g., APT activity, emerging attack vectors) that can be used to dynamically adjust the vulnerability subgraph by adding newly discovered vulnerabilities, hypothesizing the presence of zero-day vulnerabilities in seemingly secure but critical components, and adjusting severity scores accordingly. By modeling these external risk factors and mapping them to other constructs in the various subgraphs, the multi-graph model becomes context-aware and capable of evolving in response to changing environmental conditions, such as geopolitical events or new regulatory frameworks. 

\begin{figure}[htbp]
  \centering
  \includegraphics[width=.6\linewidth]{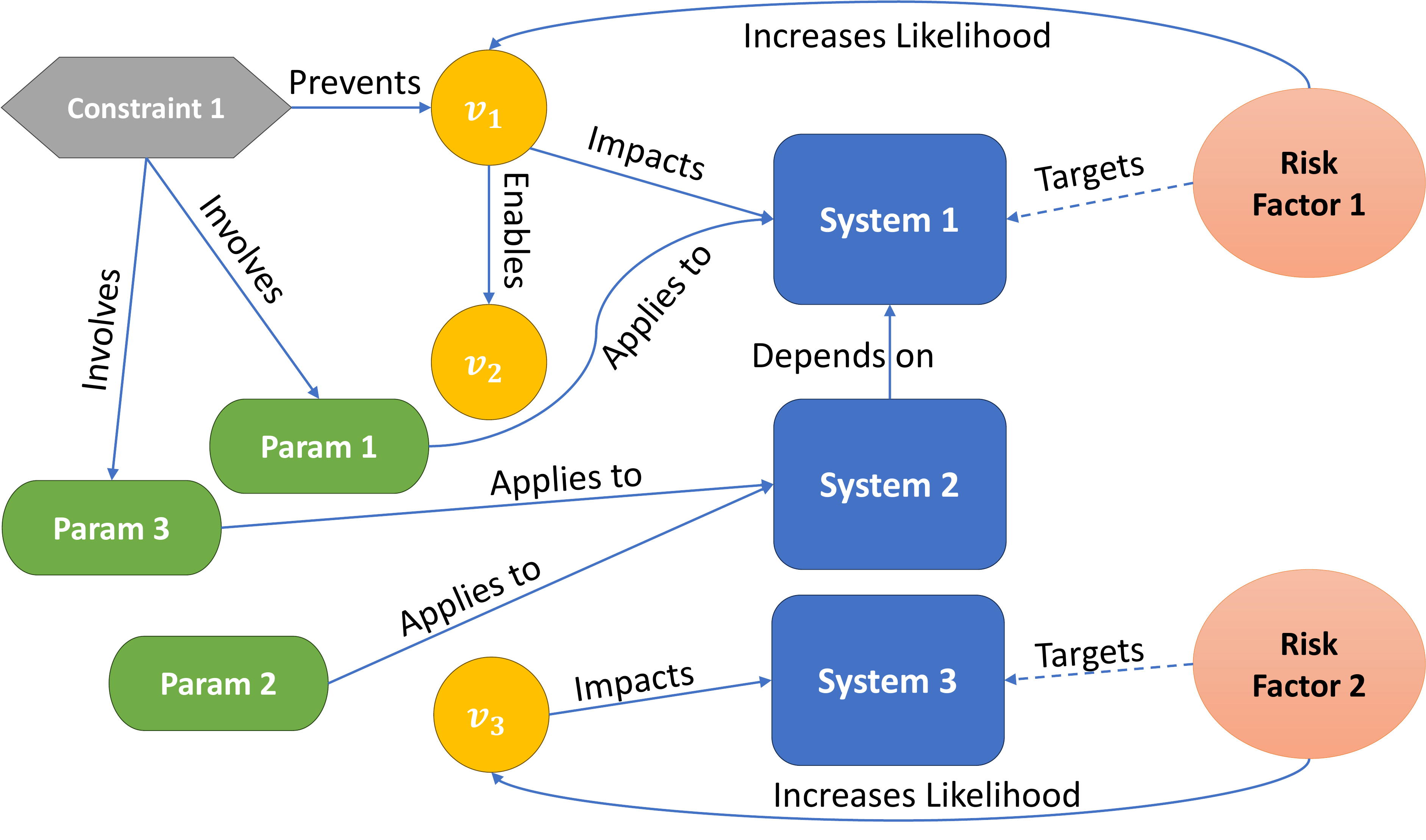}
  \caption{Example of graphical model.}
  \label{fig:graph}
\end{figure}

In our graphical data model, nodes represent key entities that capture different aspects of the system and its risk management environment, as shown in the in the notional example of \Fig~\ref{fig:graph}. Vulnerability nodes (depicted in yellow) capture weaknesses or flaws in the system that can be exploited by threats, offering insights into potential security breaches and multi-step attacks~\cite{albanese2018graphical,liu2008multistep,navarro2018multistep} --- enabled by the sequential exploitation of multiple dependent vulnerabilities --- and their impact on system components. Component nodes (in blue) represent the physical or logical parts of a system at different levels of abstraction (individual components, subsystems, or systems), with the component subgraph capturing functional dependencies among the parts. Risk Factor nodes (in orange) define variables or conditions, identified through Module 2.1, that target specific systems and contribute to the likelihood or impact of security incidents, enabling a contextual assessment of threats and system exposure to external risks. Configuration nodes (in green) document the settings and parameters that govern the operation of the system, reflecting its current state and influencing its vulnerability and resilience to attacks. Constraints among configuration parameters (in gray) capture the role of configuration settings in mitigating vulnerability exposure. Together, these interconnected sets of nodes offer a rich representation of the system’s structure, weaknesses, and risk environment, providing a baseline knowledge for LLMs to reason about.

The multi-graph model is designed to scale efficiently with large and complex systems, allowing it to handle a growing number of system components, vulnerabilities, and external factors. Scalability is key to ensuring that the framework remains effective as organizations continue to expand their cybersecurity infrastructure and adopt new tools and technologies. To this aim, the model builds on our previous work on efficiently indexing patterns in graph data~\cite{albanese2007magic}. Building on our previous work on vulnerability metrics~\cite{albanese2023framework} and quantification of the potential impact of cyber-attacks~\cite{iganibo2023attack}, we define families of metrics to quantify risk and guide the generation of courses of action that can provably reduce risk for the enterprise, paving the way for a novel Metrics-Augmented Generation (MAG). This system of metrics builds upon the two key vulnerability metrics defined in~\cite{albanese2023framework}: the exploitation likelihood $\rho(v)$ and the exposure factor $ef(v)$ of a vulnerability $v$, as defined by Equations 1 and 2 below.

\begin{equation}
\rho(v) = \frac{\prod_{x \in X_l^\uparrow} \left( 1 - e^{-\alpha_x f_x(X(v))} \right)}{\prod_{x \in X_l^\downarrow} e^{\alpha_x f_x(X(v))}}
\end{equation}

\begin{equation}
ef(v) = \frac{\prod_{x \in X_e^\uparrow} \left( 1 - e^{-\alpha_x f_x(X(v))} \right)}{\prod_{x \in X_e^\downarrow} e^{\alpha_x f_x(X(v))}}
\end{equation}

In these equations, $\mathcal{X}_l^\uparrow$, $\mathcal{X}_l^\downarrow$, $\mathcal{X}_e^\uparrow$, and $\mathcal{X}_e^\downarrow$ are sets of variables that influence the exploitation likelihood and exposure factors of vulnerabilities. Specifically, $\mathcal{X}_l^\uparrow$ and $\mathcal{X}_l^\downarrow$ are sets of variables that, respectively, contribute to increasing and decreasing the likelihood as their values increase. Similarly, $\mathcal{X}_e^\uparrow$ and $\mathcal{X}_e^\downarrow$ are sets of variables that, respectively, contribute to increasing and decreasing the exposure as their values increase.

Typical examples of variables in $\mathcal{X}_l^\uparrow$ include, but are not limited to, a vulnerability’s exploitability score, the time since details about the vulnerability were published, and the set of known exploits. In our work, we model the external risk factors discussed earlier as variables in $X_l^\uparrow$, allowing us great flexibility in modeling. Examples of variables in $\mathcal{X}_l^\downarrow$ include, but are not limited to, the set of known Intrusion Detection System (IDS) rules associated with a vulnerability and the set of available vulnerability scanning plugins. Similarly, examples of variables in $\mathcal{X}_e^\uparrow$ include a vulnerability’s impact score, and examples of variables in $\mathcal{X}_e^\downarrow$ include the set of deployed IDS rules associated with a vulnerability, which can help decrease the impact by detecting the onset of exploitation activity.

In these equations, each variable contributes to the overall likelihood or exposure factor as a multiplicative factor between 0 and 1 that is formulated to account for diminishing returns. Factors corresponding to a variable $X$ in $\mathcal{X}_l^\uparrow$ or $\mathcal{X}_e^\uparrow$ are of the form $1 - e^{-\alpha_X \cdot f_X (X(v))}$ where $\alpha_X$ is a tunable parameter, $X(v)$ is the value of $X$ for $v$, and $f_X$ is a monotonically increasing function used to convert values of $X$ to scalar values. Similarly, factors corresponding to a variable $X$ in $\mathcal{X}_l^\downarrow$ or $\mathcal{X}_e^\downarrow$ are of the form $e^{-\alpha_X \cdot f_X (X(v))}$.

A key aspect of our approach is to model and quantify the role of external risk factors. In complex enterprise networks comprising multiple subsystems, each providing distinct services, various risk factors may impact systems differently. To capture these influences, we map each risk factor to the vulnerabilities existing on its target system or subsystem and incorporate these factors into the computation of vulnerability likelihood in Equation~(1) as described earlier. In the notional example of \Fig~\ref{fig:graph}, two risk factors respectively affect two different systems, so they can be modeled as variables influencing the likelihood of exposed vulnerabilities $v_1$ and $v_3$ respectively.

%----------------------------------------------------------------------------------
\subsection{Module 3: LLM-Assisted Adaptive Decision-Making} \label{sec:module-3}
%----------------------------------------------------------------------------------

Module 3 uses LLMs to dynamically assess and adapt to the ever-changing risk landscape, and it responds the following research questions.

\begin{shadedbox} \textbf{Research Questions}.  How can SOCs
dynamically assess and adapt to the evolving risk landscape, and use
LLM-based agents to support real-time risk assessment and prioritize
incident response efforts? What are the most effective models for
human-machine collaboration in SOCs, to support decision-making and
adapt to an ever-changing threat environment?
\end{shadedbox}

%----------------------------------------------------------------------------------
\subsubsection{Module 3.1: Interrogating LLM for Context-aware Investigation}
%----------------------------------------------------------------------------------

This submodule analyzes data from diverse feeds, including network
traffic and alerts from various tools, inferences from larger contexts
(Module 2.1), and risk indications calculated from the graphical model
(Module 2.2), to suggest potential courses of actions both for
investigation and remediation. Anthropological fieldwork provides an effective means of capturing tacit analyst knowledge to develop this capability. Embedded
fieldworkers carry out participant observation~\cite{gunn2014participant, spradley2016participant} and work as
analysts in real-world SOCs.
Embedded fieldworkers play the roles of both the researcher
and the analyst shown in Fig.~\ref{fig:vision}. In carrying out the investigation,
the analysts ask the LLM agent to perform subtasks such as
looking for patterns of potential misuse in logs. The LLM is provided with all relevant contextual information from the various
data feeds described above. It generates an initial output,
including a rationale explaining its decision-making process. Analysts
then interact with the LLM by refining its outputs through
iterative natural language instructions, guiding the model to better
align with their preferences. This refinement process could involve
multiple rounds of revisions, where the LLM adjusts its reasoning and
suggestions in response to user feedback. If the analysts are unable
to achieve a satisfactory result through this interaction alone, they
could directly edit the output to produce a revised version. These
revised outputs, along with the original inputs, are then used to
fine-tune the LLM, gradually aligning its responses with analyst
preferences.

This approach builds on the concept of generative feedback~\cite{jin2023alignment}, which emphasizes learning from natural language critiques to iteratively
refine model outputs. Rather than relying on static supervision or
RLHF, generative feedback enables LLMs to adapt dynamically through
user interactions that provide fine-grained feedback on both the
strengths and weaknesses of an initial response. By incorporating
iterative refinement and direct human correction, the framework extends
this paradigm to better support investigative workflows, allowing the
LLM to capture subject matter expertise and institutional preferences
over successive interactions. In the early stages, analysts may need
to invest more effort in refining outputs, either by providing
detailed prompting or by manually editing the model’s
responses. However, as the LLM incorporates this expertise through
feedback-driven learning, the need for direct human intervention is
expected to decrease. All interactions are recorded as op traces,
as illustrated in Fig.~\ref{fig:vision}. After sufficient improvement, the agent is piloted with production SOC analysts. At this stage, the fieldworkers
only play the role of the researcher in Fig.~\ref{fig:vision}, alongside SOC
analysts, to observe how human analysts interact with the LLM agent.
In addition to the op traces, the embedded researchers also record
their observations in fieldnotes. These observations can reveal
unspoken suppositions and other dimensions of tacit knowledge not
captured in the conversations between human analysts and LLM
agents. These tacit dimensions are useful in creating more
effective learning signals for LLM agents to be improved through the
learning process described below. The continuous evaluation and
refinement aim to enhance the precision and effectiveness of
human-machine co-teaming for investigating cyber incidents and making
the right remediation decisions.

%----------------------------------------------------------------------------------
\subsubsection{Module 3.2: Teaching LLM through SOC Experience}
%----------------------------------------------------------------------------------

This submodule provides approaches that enable the LLM agents
to learn from their experience in the SOC’s
investigative tasks. The learning signals generated through LLM
agents’ collaboration with human analysts come from two sources
(Fig.~\ref{fig:vision}): 1) recorded op traces, and 2) distilled explicit knowledge
through qualitative analysis. To make the learning process
sustainable, human analysts bear minimum burden in helping
provide the learning signals. The conversations between the LLM agent
and human analysts must be organic and for the sole purpose of moving
forward with the investigation and finding the most appropriate
remediations. In doing so some tacit knowledge is inevitably
missed in the recorded op traces. This is where the embedded
researcher can help. The researcher, trained in anthropological
research methods, records in the fieldnotes any observations that
capture the context under which the interactions took place. The
analysis of the fieldnotes provide
additional learning signals from the explicated tacit knowledge
through qualitative analysis.  After the learning signals are formed,
various learning strategies are applied as explained in
Section~\ref{sec:learning_strategies}. Different strategies are compared to
understand their strengths and weaknesses and gain an
understanding of when to utilize which strategies to yield the best
outcome. After the LLM agent has learned substantial new knowledge, an
updated version is deployed into operation, and the process
repeats.

%----------------------------------------------------------------------------------
\subsection{Module 4: Explainable AI and Impact Quantification} \label{sec:module-4}
%----------------------------------------------------------------------------------
Module 4 employs LLMs to embed explainable-AI capabilities within SOC workflows, thereby improving transparency and trust in incident-response decisions and providing metrics to quantify the operational impact of AI adoption. It addresses the following research question.

\begin{shadedbox} \textbf{Research Questions}.  How can we leverage an
LLM’s ability to articulate reasoning to achieve explainable AI
integrated into SOCs to enhance transparency, trust, and understanding
of decision-making during incident response? How can the impact of
adopting LLMs in SOCs be measured, not only in terms of operational
efficiency but also in incident detection and response effectiveness?
\end{shadedbox}

%----------------------------------------------------------------------------------
\subsubsection{Module 4.1: LLM-based Report Generation}
%----------------------------------------------------------------------------------

Once human analysts, assisted by LLM agents, have reached
investigative conclusions and formulated remediation strategies, the
SOC needs to produce reports for the relevant stakeholders who are
impacted by the incident and/or remediations. 
The framework leverages LLM’s capability to transform the analysis results from
Module 3.1 into reports designed for various human stakeholders,
emphasizing interpretability and transparency. The report provides
a narrative explanation of the proposed actions, their rationale, and
their alignment with risk metrics. It also includes contextual
information, such as the expected impact, underlying assumptions, and
data sources, enabling stakeholders to understand and trust the
decision-making process. To help the AI develop potential courses of
action, a template is developed by the research team with input
from SOC team members. Each course of action presents SOC-specific details and clearly links the proposed steps to the particular security issue they address.

%----------------------------------------------------------------------------------
\subsubsection{Module 4.2: Plan Evaluation} 
%----------------------------------------------------------------------------------

In Module 4.2, the focus shifts to evaluating the proposed plan's feasibility and effectiveness
through a structured review by human analysts. This evaluation differs
from the interactive refinement in Module 3.2 by incorporating a
broader, post-generation assessment of the plan, including its
alignment with organizational goals and long-term impact. Feedback
from this evaluation refines both the analysis process in Module
3.1 and the report generation process in Module 4.1, ensuring
iterative improvement and consistency with evolving security contexts.

%----------------------------------------------------------------------------------
\subsubsection{Module 4.3: Plan Execution}
%----------------------------------------------------------------------------------

Module 4.3 addresses the automation of the plan's implementation,
transforming the finalized plan from Module 4.1 into executable
artifacts such as scripts, system configurations, or other
machine-readable instructions. This task focuses on operationalizing
the plan with minimal manual intervention while maintaining oversight
mechanisms to ensure alignment with organizational policies. By
bridging the gap between planning and action, Module 4.3 enables
efficient and reliable execution of risk mitigation strategies,
reinforcing the overall AI-assistance framework.

%% file: discussion.tex
% !TEX root = main.tex

%----------------------------------------------------------------------------------
\section{Discussion and Future Work} \label{sec:discussion}
%----------------------------------------------------------------------------------

While AI-driven co-teaming presents transformative potential, challenges remain, including trust in AI outputs, explainability, and continuous learning from SOC interactions. Future research should explore fine-tuning SOC-specific LLMs, enhancing AI explainability, and integrating human feedback loops for iterative learning.

%----------------------------------------------------------------------------------
\subsection{Representative Case Studies} 
%----------------------------------------------------------------------------------

Cyber-attacks targeting high-profile organizations are often driven by geopolitical tensions, exposing critical gaps in SOCs, particularly in integrating external threat intelligence and leveraging human-machine collaboration for increased efficiency and adaptive response to evolving threats. Two major incidents --- the 2014 Sony Pictures Entertainment attack and the 2016 DNC cyber intrusions --- illustrate these challenges.

\begin{figure}[htbp]
  \centering
  \includegraphics[width=\linewidth]{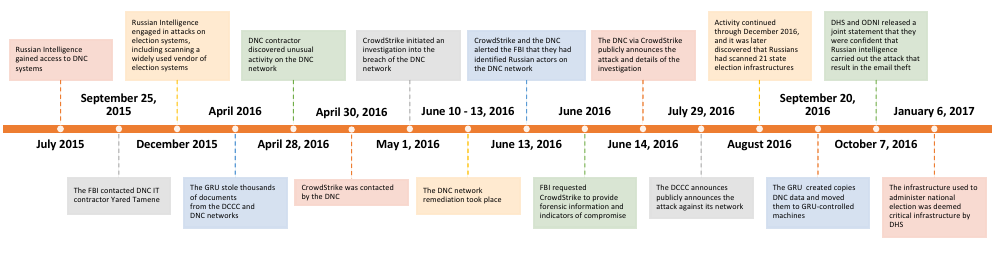}
  \caption{Timeline of the DNC attack.}
  \label{fig:dnc-attack}
\end{figure}

The Sony attack, attributed to North Korea~\cite{trendmicro2018lazarus} in retaliation for the release of the movie \emph{The Interview}~\cite{trendmicro2014sony}, began with a phishing campaign that compromised Sony's network. The attackers deployed Destover malware, wiping data and rendering systems inoperable while simultaneously exfiltrating sensitive data, including unreleased films, emails, and employee records. The breach caused severe operational and reputational damage, highlighting the failure of traditional SOCs to anticipate politically motivated cyber threats. Similarly, the DNC breaches, orchestrated by Russian state-sponsored groups Cozy Bear and Fancy Bear~\cite{crowdstrike2020dnc}, relied on spear-phishing to gain initial access. Cozy Bear persisted within the network for months before Fancy Bear executed more aggressive data exfiltration and public leaks, influencing U.S. politics (see \Fig~\ref{fig:dnc-attack}). These attacks underscored the difficulty of identifying and responding to prolonged, stealthy threats in a complex geopolitical landscape.

Our framework would have significantly enhanced a SOC's ability to detect and mitigate such attacks. By incorporating geopolitical threat signals into SOC workflows, harmonizing disparate data sources, and leveraging LLM-based AI for real-time contextual analysis, our approach would have helped provide analysts with some actionable intelligence well before the attacks escalated. Rather than relying on static detection mechanisms, our adaptive AI-driven co-teaming paradigm would have surfaced geopolitical risks, linked anomalous activity to external threat actors, and generated explainable remediation plans, allowing human analysts to act swiftly and effectively. The Sony and DNC incidents illustrate the urgent need for a human-machine collaboration model that dynamically integrates external intelligence, anticipates emerging threats, and enables proactive defense.

%% file: conclusions.tex
% !TEX root = main.tex

%----------------------------------------------------------------------------------
\section{Conclusions} \label{sec:conclusions}
%----------------------------------------------------------------------------------

This paper outlines a vision for AI-driven human-machine co-teaming in Security Operations Centers (SOCs), positioning large language models (LLMs) as collaborative apprentices that learn from and work alongside human analysts. We propose a paradigm in which LLM-based AI agents augment SOC workflows by internalizing tacit knowledge, supporting dynamic decision-making, and alleviating analyst burden without replacing human judgment. By reframing AI not as a deterministic tool but as a learning partner, we open new pathways for adaptive, context-aware SOC operations. We invite the research and practitioner community to explore this human-AI collaboration model, investigate its practical applications, and advance its development toward measurable improvements in SOC productivity and resilience.

%% file: ack.tex
\section*{Acknowledgment}

This work was
partially supported by the National Science
Foundation under award no. 2235102, Office of Naval Research under
award no. N00014-23-1-2538, and National Security Agency under award No. H98230-22-1-0311.
Any opinions, findings and
conclusions or recommendations expressed in this material are those of
the authors and do not necessarily reflect the views of the above funding agencies.

%%% Local Variables:
%%% mode: latex
%%% TeX-master: "main"
%%% End:

%% file: main.bbl

\begin{thebibliography}{60}

%%% the Web address.

\ifx \showCODEN    \undefined \def \showCODEN     #1{\unskip}     \fi
\ifx \showDOI      \undefined \def \showDOI       #1{#1}\fi
\ifx \showISBNx    \undefined \def \showISBNx     #1{\unskip}     \fi
\ifx \showISBNxiii \undefined \def \showISBNxiii  #1{\unskip}     \fi
\ifx \showISSN     \undefined \def \showISSN      #1{\unskip}     \fi
\ifx \showLCCN     \undefined \def \showLCCN      #1{\unskip}     \fi
\ifx \shownote     \undefined \def \shownote      #1{#1}          \fi
\ifx \showarticletitle \undefined \def \showarticletitle #1{#1}   \fi
\ifx \showURL      \undefined \def \showURL       {\relax}        \fi
% invisible to TeX
\providecommand\bibfield[2]{#2}
\providecommand\bibinfo[2]{#2}
\providecommand\natexlab[1]{#1}
\providecommand\showeprint[2][]{arXiv:#2}

\bibitem[\protect\citeauthoryear{Akhgari, Malekimajd, and Rahmani}{Akhgari
  et~al\mbox{.}}{2022}]%
        {akhgari2022sem}
\bibfield{author}{\bibinfo{person}{Z. Akhgari}, \bibinfo{person}{M.
  Malekimajd}, {and} \bibinfo{person}{H. Rahmani}.}
  \bibinfo{year}{2022}\natexlab{}.
\newblock \showarticletitle{Sem-TED: Semantic Twitter Event Detection and
  Adapting with News Stories}. In \bibinfo{booktitle}{\emph{Proceedings of the
  International Conference on Web Research}}. \bibinfo{pages}{61--69}.
\newblock
\urldef\tempurl%
\url{https://doi.org/10.1109/ICWR54782.2022.9786234}
\showDOI{\tempurl}


\bibitem[\protect\citeauthoryear{Al-Mhiqani, Ahmad, Abidin, Yassin, Hassan,
  Abdulkareem, Ali, and Yunos}{Al-Mhiqani et~al\mbox{.}}{2020}]%
        {almhiqani2020insider}
\bibfield{author}{\bibinfo{person}{M.~N. Al-Mhiqani}, \bibinfo{person}{R.
  Ahmad}, \bibinfo{person}{Z.~Z. Abidin}, \bibinfo{person}{W. Yassin},
  \bibinfo{person}{A. Hassan}, \bibinfo{person}{K.~H. Abdulkareem},
  \bibinfo{person}{N.~S. Ali}, {and} \bibinfo{person}{Z. Yunos}.}
  \bibinfo{year}{2020}\natexlab{}.
\newblock \showarticletitle{A Review of Insider Threat Detection:
  Classification, Machine Learning Techniques, Datasets, Open Challenges, and
  Recommendations}.
\newblock \bibinfo{journal}{\emph{Applied Sciences}} \bibinfo{volume}{10},
  \bibinfo{number}{15} (\bibinfo{year}{2020}), \bibinfo{pages}{5208}.
\newblock
\urldef\tempurl%
\url{https://doi.org/10.3390/app10155208}
\showDOI{\tempurl}


\bibitem[\protect\citeauthoryear{Alahmadi, Axon, and Martinovic}{Alahmadi
  et~al\mbox{.}}{2022}]%
        {alahmadi2022false}
\bibfield{author}{\bibinfo{person}{B.~A. Alahmadi}, \bibinfo{person}{L. Axon},
  {and} \bibinfo{person}{I. Martinovic}.} \bibinfo{year}{2022}\natexlab{}.
\newblock \showarticletitle{99\% False Positives: A Qualitative Study of SOC
  Analysts’ Perspectives on Security Alarms}. In
  \bibinfo{booktitle}{\emph{Proceedings of the 31st USENIX Security Symposium
  (USENIX Security 2022)}}. \bibinfo{address}{Boston, MA, USA},
  \bibinfo{pages}{2783--2800}.
\newblock
\urldef\tempurl%
\url{https://www.usenix.org/system/files/sec22-alahmadi.pdf}
\showURL{%
\tempurl}


\bibitem[\protect\citeauthoryear{Albanese, Iganibo, and Adebiyi}{Albanese
  et~al\mbox{.}}{2023}]%
        {albanese2023framework}
\bibfield{author}{\bibinfo{person}{M. Albanese}, \bibinfo{person}{I. Iganibo},
  {and} \bibinfo{person}{O. Adebiyi}.} \bibinfo{year}{2023}\natexlab{}.
\newblock \showarticletitle{A Framework for Designing Vulnerability Metrics}.
\newblock \bibinfo{journal}{\emph{Computers \& Security}}
  \bibinfo{volume}{132} (\bibinfo{date}{September} \bibinfo{year}{2023}),
  \bibinfo{pages}{11 pages}.
\newblock
\urldef\tempurl%
\url{https://doi.org/10.1016/j.cose.2023.103382}
\showDOI{\tempurl}


\bibitem[\protect\citeauthoryear{Albanese and Jajodia}{Albanese and
  Jajodia}{2018}]%
        {albanese2018graphical}
\bibfield{author}{\bibinfo{person}{M. Albanese} {and} \bibinfo{person}{S.
  Jajodia}.} \bibinfo{year}{2018}\natexlab{}.
\newblock \showarticletitle{A Graphical Model to Assess the Impact of
  Multi-Step Attacks}.
\newblock \bibinfo{journal}{\emph{The Journal of Defense Modeling and
  Simulation}} \bibinfo{volume}{15}, \bibinfo{number}{1}
  (\bibinfo{year}{2018}), \bibinfo{pages}{79--93}.
\newblock
\urldef\tempurl%
\url{https://doi.org/10.1177/1548512917706043}
\showDOI{\tempurl}


\bibitem[\protect\citeauthoryear{Albanese, Pugliese, Subrahmanian, and
  Udrea}{Albanese et~al\mbox{.}}{2007}]%
        {albanese2007magic}
\bibfield{author}{\bibinfo{person}{M. Albanese}, \bibinfo{person}{A. Pugliese},
  \bibinfo{person}{V.~S. Subrahmanian}, {and} \bibinfo{person}{O. Udrea}.}
  \bibinfo{year}{2007}\natexlab{}.
\newblock \showarticletitle{MAGIC: A Multi-Activity Graph Index for Activity
  Detection}. In \bibinfo{booktitle}{\emph{Proceedings of the 2007 IEEE
  International Conference on Information Reuse and Integration (IRI 2007)}}.
  \bibinfo{address}{Las Vegas, NV, USA}, \bibinfo{pages}{267--272}.
\newblock
\urldef\tempurl%
\url{https://doi.org/10.1109/IRI.2007.4296632}
\showDOI{\tempurl}


\bibitem[\protect\citeauthoryear{Allan, Carbonell, Doddington, Yamron, and
  Yang}{Allan et~al\mbox{.}}{1998}]%
        {allan1998topic}
\bibfield{author}{\bibinfo{person}{J. Allan}, \bibinfo{person}{J. Carbonell},
  \bibinfo{person}{G. Doddington}, \bibinfo{person}{J. Yamron}, {and}
  \bibinfo{person}{Y. Yang}.} \bibinfo{year}{1998}\natexlab{}.
\newblock \showarticletitle{Topic Detection and Tracking Pilot Study: Final
  Report}. In \bibinfo{booktitle}{\emph{Proceedings of the DARPA Broadcast News
  Transcription and Understanding Workshop}}. \bibinfo{pages}{194--218}.
\newblock
\urldef\tempurl%
\url{https://ciir.cs.umass.edu/pubfiles/ir-137.pdf}
\showURL{%
\tempurl}


\bibitem[\protect\citeauthoryear{Almazrouei, Alobeidli, Alshamsi, Cappelli,
  Cojocaru, Debbah, Goffinet, Hesslow, Launay, and Malartic}{Almazrouei
  et~al\mbox{.}}{2023}]%
        {almazrouei2023falcon}
\bibfield{author}{\bibinfo{person}{E. Almazrouei}, \bibinfo{person}{H.
  Alobeidli}, \bibinfo{person}{A. Alshamsi}, \bibinfo{person}{A. Cappelli},
  \bibinfo{person}{R. Cojocaru}, \bibinfo{person}{M. Debbah},
  \bibinfo{person}{É. Goffinet}, \bibinfo{person}{D. Hesslow},
  \bibinfo{person}{J. Launay}, {and} \bibinfo{person}{Q. Malartic}.}
  \bibinfo{year}{2023}\natexlab{}.
\newblock \showarticletitle{The Falcon Series of Open Language Models}.
\newblock \bibinfo{journal}{\emph{arXiv}} (\bibinfo{year}{2023}).
\newblock
\urldef\tempurl%
\url{https://doi.org/10.48550/arXiv.2311.16867}
\showDOI{\tempurl}


\bibitem[\protect\citeauthoryear{Archive}{Archive}{[n.d.]}]%
        {internetarchive2025wayback}
\bibfield{author}{\bibinfo{person}{Internet Archive}.}
  \bibinfo{year}{[n.d.]}\natexlab{}.
\newblock \bibinfo{title}{Wayback Machine}.
\newblock
\newblock
\urldef\tempurl%
\url{http://web.archive.org/}
\showURL{%
\tempurl}
\newblock
\shownote{[Accessed: Jan. 24, 2025].}


\bibitem[\protect\citeauthoryear{Asgari-Chenaghlu, Feizi-Derakhshi, Farzinvash,
  Balafar, and Motamed}{Asgari-Chenaghlu et~al\mbox{.}}{2021}]%
        {asgari2021topic}
\bibfield{author}{\bibinfo{person}{M. Asgari-Chenaghlu}, \bibinfo{person}{M.-R.
  Feizi-Derakhshi}, \bibinfo{person}{L. Farzinvash}, \bibinfo{person}{M.-A.
  Balafar}, {and} \bibinfo{person}{C. Motamed}.}
  \bibinfo{year}{2021}\natexlab{}.
\newblock \showarticletitle{Topic Detection and Tracking Techniques on Twitter:
  A Systematic Review}.
\newblock \bibinfo{journal}{\emph{Complexity}}  \bibinfo{volume}{2021}
  (\bibinfo{year}{2021}), \bibinfo{pages}{Article ID 8833084}.
\newblock
\urldef\tempurl%
\url{https://doi.org/10.1155/2021/8833084}
\showDOI{\tempurl}


\bibitem[\protect\citeauthoryear{Barnum}{Barnum}{2012}]%
        {barnum2012stix}
\bibfield{author}{\bibinfo{person}{S. Barnum}.}
  \bibinfo{year}{2012}\natexlab{}.
\newblock \bibinfo{title}{Standardizing Cyber Threat Intelligence Information
  with the Structured Threat Information Expression (STIX\texttrademark)}.
\newblock \bibinfo{howpublished}{The MITRE Corporation}.
\newblock
\urldef\tempurl%
\url{https://www.mitre.org/sites/default/files/publications/stix.pdf}
\showURL{%
\tempurl}


\bibitem[\protect\citeauthoryear{Bayer, Kuehn, Shanehsaz, and Reuter}{Bayer
  et~al\mbox{.}}{2024}]%
        {bayer2024cysecbert}
\bibfield{author}{\bibinfo{person}{M. Bayer}, \bibinfo{person}{P. Kuehn},
  \bibinfo{person}{R. Shanehsaz}, {and} \bibinfo{person}{C. Reuter}.}
  \bibinfo{year}{2024}\natexlab{}.
\newblock \showarticletitle{CySecBERT: A Domain-Adapted Language Model for the
  Cybersecurity Domain}.
\newblock \bibinfo{journal}{\emph{ACM Transactions on Privacy and Security}}
  \bibinfo{volume}{27}, \bibinfo{number}{2} (\bibinfo{year}{2024}),
  \bibinfo{pages}{1--20}.
\newblock
\urldef\tempurl%
\url{https://doi.org/10.1145/3652594}
\showDOI{\tempurl}


\bibitem[\protect\citeauthoryear{Bernard, Wutich, and Ryan}{Bernard
  et~al\mbox{.}}{2016}]%
        {bernard2016analyzing}
\bibfield{author}{\bibinfo{person}{H.~R. Bernard}, \bibinfo{person}{A. Wutich},
  {and} \bibinfo{person}{G.~W. Ryan}.} \bibinfo{year}{2016}\natexlab{}.
\newblock \bibinfo{booktitle}{\emph{Analyzing Qualitative Data: Systematic
  Approaches}}.
\newblock \bibinfo{publisher}{SAGE Publications}.
\newblock
\urldef\tempurl%
\url{https://uk.sagepub.com/en-gb/eur/analyzing-qualitative-data/book240717}
\showURL{%
\tempurl}


\bibitem[\protect\citeauthoryear{Brin and Page}{Brin and Page}{1998}]%
        {brin1998search}
\bibfield{author}{\bibinfo{person}{S. Brin} {and} \bibinfo{person}{L. Page}.}
  \bibinfo{year}{1998}\natexlab{}.
\newblock \showarticletitle{The Anatomy of a Large-Scale Hypertextual Web
  Search Engine}.
\newblock \bibinfo{journal}{\emph{Computer Networks and ISDN Systems}}
  \bibinfo{volume}{30} (\bibinfo{year}{1998}), \bibinfo{pages}{107--117}.
\newblock
\urldef\tempurl%
\url{https://www.sciencedirect.com/science/article/abs/pii/S016975529800110X}
\showURL{%
\tempurl}


\bibitem[\protect\citeauthoryear{Cantini and Marozzo}{Cantini and
  Marozzo}{2022}]%
        {cantini2022topic}
\bibfield{author}{\bibinfo{person}{R. Cantini} {and} \bibinfo{person}{F.
  Marozzo}.} \bibinfo{year}{2022}\natexlab{}.
\newblock \showarticletitle{Topic Detection and Tracking in Social Media
  Platforms}. In \bibinfo{booktitle}{\emph{Proceedings of the 1st International
  Conference on Pervasive Knowledge and Collective Intelligence on Web and
  Social Media}}. \bibinfo{publisher}{Springer}, \bibinfo{pages}{41--56}.
\newblock
\urldef\tempurl%
\url{https://doi.org/10.1007/978-3-031-31469-8_3}
\showDOI{\tempurl}


\bibitem[\protect\citeauthoryear{Chen, Cui, Wang, Cao, Yang, Jiang, Lu, and
  Liu}{Chen et~al\mbox{.}}{2024}]%
        {chen2024survey}
\bibfield{author}{\bibinfo{person}{Y. Chen}, \bibinfo{person}{M. Cui},
  \bibinfo{person}{D. Wang}, \bibinfo{person}{Y. Cao}, \bibinfo{person}{P.
  Yang}, \bibinfo{person}{B. Jiang}, \bibinfo{person}{Z. Lu}, {and}
  \bibinfo{person}{B. Liu}.} \bibinfo{year}{2024}\natexlab{}.
\newblock \showarticletitle{A Survey of Large Language Models for Cyber Threat
  Detection}.
\newblock \bibinfo{journal}{\emph{Computers \& Security}} (\bibinfo{date}{July}
  \bibinfo{year}{2024}).
\newblock
\urldef\tempurl%
\url{https://doi.org/10.1016/j.cose.2024.104016}
\showDOI{\tempurl}


\bibitem[\protect\citeauthoryear{Chen, Ding, Alowain, Chen, and Wagner}{Chen
  et~al\mbox{.}}{2023}]%
        {chen2023diversevul}
\bibfield{author}{\bibinfo{person}{Y. Chen}, \bibinfo{person}{Z. Ding},
  \bibinfo{person}{L. Alowain}, \bibinfo{person}{X. Chen}, {and}
  \bibinfo{person}{D. Wagner}.} \bibinfo{year}{2023}\natexlab{}.
\newblock \showarticletitle{DiverseVul: A New Vulnerable Source Code Dataset
  for Deep Learning-Based Vulnerability Detection}. In
  \bibinfo{booktitle}{\emph{Proceedings of the 26th International Symposium on
  Research in Attacks, Intrusions, and Defenses (RAID 2023)}}.
  \bibinfo{pages}{654--668}.
\newblock
\urldef\tempurl%
\url{https://doi.org/10.1145/3607199.3607242}
\showDOI{\tempurl}


\bibitem[\protect\citeauthoryear{CrowdStrike}{CrowdStrike}{2020}]%
        {crowdstrike2020dnc}
\bibfield{author}{\bibinfo{person}{CrowdStrike}.}
  \bibinfo{year}{2020}\natexlab{}.
\newblock \bibinfo{title}{CrowdStrike’s Work with the Democratic National
  Committee: Setting the Record Straight}.
\newblock \bibinfo{howpublished}{CrowdStrike Blog}.
\newblock
\urldef\tempurl%
\url{https://www.crowdstrike.com/blog/bears-midst-intrusion-democratic-national-committee/}
\showURL{%
\tempurl}


\bibitem[\protect\citeauthoryear{Deng, Liu, Mayoral-Vilches, Liu, Li, Xu,
  Zhang, Liu, Pinzger, and Rass}{Deng et~al\mbox{.}}{2024}]%
        {deng2024pentestgpt}
\bibfield{author}{\bibinfo{person}{G. Deng}, \bibinfo{person}{Y. Liu},
  \bibinfo{person}{V. Mayoral-Vilches}, \bibinfo{person}{P. Liu},
  \bibinfo{person}{Y. Li}, \bibinfo{person}{Y. Xu}, \bibinfo{person}{T. Zhang},
  \bibinfo{person}{Y. Liu}, \bibinfo{person}{M. Pinzger}, {and}
  \bibinfo{person}{S. Rass}.} \bibinfo{year}{2024}\natexlab{}.
\newblock \showarticletitle{PentestGPT: Evaluating and Harnessing Large
  Language Models for Automated Penetration Testing}. In
  \bibinfo{booktitle}{\emph{Proceedings of the 33rd USENIX Security Symposium
  (USENIX Security 24)}}. \bibinfo{address}{Philadelphia, PA, USA},
  \bibinfo{pages}{847--864}.
\newblock
\urldef\tempurl%
\url{https://www.usenix.org/conference/usenixsecurity24/presentation/deng}
\showURL{%
\tempurl}


\bibitem[\protect\citeauthoryear{Fan, Ding, Ning, Wang, Li, Yin, Chua, and
  Li}{Fan et~al\mbox{.}}{2024}]%
        {fan2024rag}
\bibfield{author}{\bibinfo{person}{W. Fan}, \bibinfo{person}{Y. Ding},
  \bibinfo{person}{L. Ning}, \bibinfo{person}{S. Wang}, \bibinfo{person}{H.
  Li}, \bibinfo{person}{D. Yin}, \bibinfo{person}{T.-S. Chua}, {and}
  \bibinfo{person}{Q. Li}.} \bibinfo{year}{2024}\natexlab{}.
\newblock \showarticletitle{A Survey on RAG Meeting LLMs: Towards
  Retrieval-Augmented Large Language Models}. In
  \bibinfo{booktitle}{\emph{Proceedings of the 30th ACM SIGKDD International
  Conference on Knowledge Discovery and Data Mining}}.
  \bibinfo{address}{Barcelona, Spain}, \bibinfo{pages}{6491--6501}.
\newblock
\urldef\tempurl%
\url{https://doi.org/10.1145/3637528.3671470}
\showDOI{\tempurl}


\bibitem[\protect\citeauthoryear{Fetters and Rubinstein}{Fetters and
  Rubinstein}{2019}]%
        {fetters2019content}
\bibfield{author}{\bibinfo{person}{M.~D. Fetters} {and} \bibinfo{person}{E.~B.
  Rubinstein}.} \bibinfo{year}{2019}\natexlab{}.
\newblock \showarticletitle{The 3 Cs of Content, Context, and Concepts: A
  Practical Approach to Recording Unstructured Field Observations}.
\newblock \bibinfo{journal}{\emph{The Annals of Family Medicine}}
  \bibinfo{volume}{17}, \bibinfo{number}{6} (\bibinfo{year}{2019}),
  \bibinfo{pages}{554--560}.
\newblock
\urldef\tempurl%
\url{https://pmc.ncbi.nlm.nih.gov/articles/PMC6846267/}
\showURL{%
\tempurl}


\bibitem[\protect\citeauthoryear{for Strategic and (CSIS)}{for Strategic and
  (CSIS)}{[n.d.]}]%
        {csis2025cyberincidents}
\bibfield{author}{\bibinfo{person}{Center for Strategic} {and}
  \bibinfo{person}{International~Studies (CSIS)}.}
  \bibinfo{year}{[n.d.]}\natexlab{}.
\newblock \bibinfo{title}{Significant Cyber Incidents}.
\newblock \bibinfo{howpublished}{Strategic Technologies Program}.
\newblock
\urldef\tempurl%
\url{https://www.csis.org/programs/strategic-technologies-program/significant-cyber-incidents}
\showURL{%
\tempurl}
\newblock
\shownote{[Accessed: Jan. 24, 2025].}


\bibitem[\protect\citeauthoryear{Georgescu}{Georgescu}{2020}]%
        {georgescu2020nlpcyber}
\bibfield{author}{\bibinfo{person}{T.-M. Georgescu}.}
  \bibinfo{year}{2020}\natexlab{}.
\newblock \showarticletitle{Natural Language Processing Model for Automatic
  Analysis of Cybersecurity-Related Documents}.
\newblock \bibinfo{journal}{\emph{Symmetry}} \bibinfo{volume}{12},
  \bibinfo{number}{3} (\bibinfo{year}{2020}), \bibinfo{pages}{354}.
\newblock
\urldef\tempurl%
\url{https://doi.org/10.3390/sym12030354}
\showDOI{\tempurl}


\bibitem[\protect\citeauthoryear{Gunn and Løgstrup}{Gunn and
  Løgstrup}{2014}]%
        {gunn2014participant}
\bibfield{author}{\bibinfo{person}{W. Gunn} {and} \bibinfo{person}{L.~B.
  Løgstrup}.} \bibinfo{year}{2014}\natexlab{}.
\newblock \showarticletitle{Participant Observation, Anthropology Methodology
  and Design Anthropology Research Inquiry}.
\newblock \bibinfo{journal}{\emph{Arts and Humanities in Higher Education}}
  \bibinfo{volume}{13}, \bibinfo{number}{4} (\bibinfo{year}{2014}),
  \bibinfo{pages}{428--442}.
\newblock
\urldef\tempurl%
\url{https://doi.org/10.1177/1474022214543874}
\showDOI{\tempurl}


\bibitem[\protect\citeauthoryear{Happe and Cito}{Happe and Cito}{2023}]%
        {happe2023penetration}
\bibfield{author}{\bibinfo{person}{A. Happe} {and} \bibinfo{person}{J. Cito}.}
  \bibinfo{year}{2023}\natexlab{}.
\newblock \showarticletitle{Getting Pwn’d by AI: Penetration Testing with
  Large Language Models}. In \bibinfo{booktitle}{\emph{Proceedings of the 31st
  ACM Joint European Software Engineering Conference and Symposium on the
  Foundations of Software Engineering (ESEC/FSE 2023)}}. \bibinfo{address}{San
  Francisco, CA, USA}, \bibinfo{pages}{2082--2086}.
\newblock
\urldef\tempurl%
\url{https://doi.org/10.1145/3611643.3613083}
\showDOI{\tempurl}


\bibitem[\protect\citeauthoryear{Hasanov, Virtanen, Hakkala, and
  Isoaho}{Hasanov et~al\mbox{.}}{2024}]%
        {hasanov2024llm}
\bibfield{author}{\bibinfo{person}{I. Hasanov}, \bibinfo{person}{S. Virtanen},
  \bibinfo{person}{A. Hakkala}, {and} \bibinfo{person}{J. Isoaho}.}
  \bibinfo{year}{2024}\natexlab{}.
\newblock \showarticletitle{Application of Large Language Models in
  Cybersecurity: A Systematic Literature Review}.
\newblock \bibinfo{journal}{\emph{IEEE Access}}  \bibinfo{volume}{12}
  (\bibinfo{year}{2024}), \bibinfo{pages}{176751--176778}.
\newblock
\urldef\tempurl%
\url{https://doi.org/10.1109/access.2024.3505983}
\showDOI{\tempurl}


\bibitem[\protect\citeauthoryear{Hays and White}{Hays and White}{2024}]%
        {hays2024incident}
\bibfield{author}{\bibinfo{person}{S. Hays} {and} \bibinfo{person}{D.~J.
  White}.} \bibinfo{year}{2024}\natexlab{}.
\newblock \showarticletitle{Employing LLMs for Incident Response Planning and
  Review}.
\newblock \bibinfo{journal}{\emph{arXiv}} (\bibinfo{date}{March}
  \bibinfo{year}{2024}).
\newblock
\urldef\tempurl%
\url{https://doi.org/10.48550/arxiv.2403.01271}
\showDOI{\tempurl}


\bibitem[\protect\citeauthoryear{Iganibo, Albanese, Mosko, Bier, and
  Brito}{Iganibo et~al\mbox{.}}{2023}]%
        {iganibo2023attack}
\bibfield{author}{\bibinfo{person}{I. Iganibo}, \bibinfo{person}{M. Albanese},
  \bibinfo{person}{M. Mosko}, \bibinfo{person}{E. Bier}, {and}
  \bibinfo{person}{A.~E. Brito}.} \bibinfo{year}{2023}\natexlab{}.
\newblock \showarticletitle{An Attack Volume Metric}.
\newblock \bibinfo{journal}{\emph{Security and Privacy}} \bibinfo{volume}{6},
  \bibinfo{number}{4} (\bibinfo{date}{July} \bibinfo{year}{2023}),
  \bibinfo{pages}{23 pages}.
\newblock
\urldef\tempurl%
\url{https://doi.org/10.1002/spy2.298}
\showDOI{\tempurl}


\bibitem[\protect\citeauthoryear{Jiang, Sablayrolles, Mensch, Bamford, Chaplot,
  de~las Casas, Bressand, Lengyel, Lample, and Saulnier}{Jiang
  et~al\mbox{.}}{2023}]%
        {jiang2023mistral}
\bibfield{author}{\bibinfo{person}{A.~Q. Jiang}, \bibinfo{person}{A.
  Sablayrolles}, \bibinfo{person}{A. Mensch}, \bibinfo{person}{C. Bamford},
  \bibinfo{person}{D.~S. Chaplot}, \bibinfo{person}{D. de~las Casas},
  \bibinfo{person}{F. Bressand}, \bibinfo{person}{G. Lengyel},
  \bibinfo{person}{G. Lample}, {and} \bibinfo{person}{L. Saulnier}.}
  \bibinfo{year}{2023}\natexlab{}.
\newblock \showarticletitle{Mistral 7B}.
\newblock \bibinfo{journal}{\emph{arXiv}} (\bibinfo{year}{2023}).
\newblock
\urldef\tempurl%
\url{https://doi.org/10.48550/arXiv.2310.06825}
\showDOI{\tempurl}


\bibitem[\protect\citeauthoryear{Jin, Mehri, Hazarika, Padmakumar, Lee, Liu,
  and Namazifar}{Jin et~al\mbox{.}}{2023}]%
        {jin2023alignment}
\bibfield{author}{\bibinfo{person}{D. Jin}, \bibinfo{person}{S. Mehri},
  \bibinfo{person}{D. Hazarika}, \bibinfo{person}{A. Padmakumar},
  \bibinfo{person}{S. Lee}, \bibinfo{person}{Y. Liu}, {and} \bibinfo{person}{M.
  Namazifar}.} \bibinfo{year}{2023}\natexlab{}.
\newblock \bibinfo{title}{Data-Efficient Alignment of Large Language Models
  with Human Feedback Through Natural Language}.
\newblock \bibinfo{howpublished}{Amazon Science}.
\newblock
\urldef\tempurl%
\url{https://www.amazon.science/publications/data-efficient-alignment-of-large-language-models-with-human-feedback-through-natural-language}
\showURL{%
\tempurl}


\bibitem[\protect\citeauthoryear{Jones, Brucker-Hahn, Fidler, and Bardas}{Jones
  et~al\mbox{.}}{2023}]%
        {jones2023work}
\bibfield{author}{\bibinfo{person}{K.~R. Jones}, \bibinfo{person}{D.~A.
  Brucker-Hahn}, \bibinfo{person}{B. Fidler}, {and} \bibinfo{person}{A.~G.
  Bardas}.} \bibinfo{year}{2023}\natexlab{}.
\newblock \showarticletitle{Work-From-Home and COVID-19: Trajectories of
  Endpoint Security Management in a Security Operations Center}. In
  \bibinfo{booktitle}{\emph{Proceedings of the 32nd USENIX Security Symposium
  (USENIX Security 2023)}}. \bibinfo{address}{Anaheim, CA, USA},
  \bibinfo{pages}{2293--2310}.
\newblock
\urldef\tempurl%
\url{https://www.usenix.org/system/files/usenixsecurity23-jones.pdf}
\showURL{%
\tempurl}


\bibitem[\protect\citeauthoryear{Jong}{Jong}{2015}]%
        {jong2015interview}
\bibfield{author}{\bibinfo{person}{S. Jong}.} \bibinfo{year}{2015}\natexlab{}.
\newblock \showarticletitle{The Interview as Anti-North Korean Propaganda}.
\newblock \bibinfo{journal}{\emph{NK News}} (\bibinfo{date}{March}
  \bibinfo{year}{2015}).
\newblock
\urldef\tempurl%
\url{https://www.nknews.org/2015/03/the-interview-as-anti-north-korean-propaganda/}
\showURL{%
\tempurl}


\bibitem[\protect\citeauthoryear{Knight}{Knight}{2023}]%
        {knight2023hinton}
\bibfield{author}{\bibinfo{person}{W. Knight}.}
  \bibinfo{year}{2023}\natexlab{}.
\newblock \showarticletitle{Geoffrey Hinton tells us why he’s now scared of
  the tech he helped build}.
\newblock \bibinfo{journal}{\emph{MIT Technology Review}} (\bibinfo{date}{May}
  \bibinfo{year}{2023}).
\newblock
\urldef\tempurl%
\url{https://www.technologyreview.com/2023/05/02/1072528/geoffrey-hinton-google-why-scared-ai/}
\showURL{%
\tempurl}


\bibitem[\protect\citeauthoryear{Kumar, Sinha, Das, Pandey, and Goswami}{Kumar
  et~al\mbox{.}}{2020}]%
        {kumar2020intrusion}
\bibfield{author}{\bibinfo{person}{V. Kumar}, \bibinfo{person}{D. Sinha},
  \bibinfo{person}{A.~K. Das}, \bibinfo{person}{S.~C. Pandey}, {and}
  \bibinfo{person}{R.~T. Goswami}.} \bibinfo{year}{2020}\natexlab{}.
\newblock \showarticletitle{An Integrated Rule-Based Intrusion Detection
  System: Analysis on UNSW-NB15 Dataset and the Real-Time Online Dataset}.
\newblock \bibinfo{journal}{\emph{Cluster Computing}}  \bibinfo{volume}{23}
  (\bibinfo{year}{2020}), \bibinfo{pages}{1397--1418}.
\newblock
\urldef\tempurl%
\url{https://doi.org/10.1007/s10586-019-03008-x}
\showDOI{\tempurl}


\bibitem[\protect\citeauthoryear{Lende, Monkhouse, Ligatti, and Ou}{Lende
  et~al\mbox{.}}{2023}]%
        {lende2023cocreation}
\bibfield{author}{\bibinfo{person}{D. Lende}, \bibinfo{person}{A. Monkhouse},
  \bibinfo{person}{J. Ligatti}, {and} \bibinfo{person}{X. Ou}.}
  \bibinfo{year}{2023}\natexlab{}.
\newblock \showarticletitle{Co-Creation in Secure Software Development: Applied
  Ethnography and the Interface of Software and Development}.
\newblock \bibinfo{journal}{\emph{Human Organization}} \bibinfo{volume}{82},
  \bibinfo{number}{1} (\bibinfo{year}{2023}), \bibinfo{pages}{13--24}.
\newblock
\urldef\tempurl%
\url{https://doi.org/10.17730/1938-3525-82.1.13}
\showDOI{\tempurl}


\bibitem[\protect\citeauthoryear{Levi, Alluouche, Ohayon, and Puzanov}{Levi
  et~al\mbox{.}}{2024}]%
        {levi2024cyberpal}
\bibfield{author}{\bibinfo{person}{M. Levi}, \bibinfo{person}{Y. Alluouche},
  \bibinfo{person}{D. Ohayon}, {and} \bibinfo{person}{A. Puzanov}.}
  \bibinfo{year}{2024}\natexlab{}.
\newblock \showarticletitle{CyberPal.AI: Empowering LLMs with Expert-Driven
  Cybersecurity Instructions}.
\newblock \bibinfo{journal}{\emph{arXiv}} (\bibinfo{date}{August}
  \bibinfo{year}{2024}).
\newblock
\urldef\tempurl%
\url{https://doi.org/10.48550/arXiv.2408.09304}
\showDOI{\tempurl}


\bibitem[\protect\citeauthoryear{Liu}{Liu}{2024}]%
        {liu2024plm}
\bibfield{author}{\bibinfo{person}{Z. Liu}.} \bibinfo{year}{2024}\natexlab{}.
\newblock \showarticletitle{A Review of Advancements and Applications of
  Pre-Trained Language Models in Cybersecurity}. In
  \bibinfo{booktitle}{\emph{Proceedings of the International Symposium on
  Digital Forensics and Security}}. \bibinfo{address}{San Antonio, TX, USA}.
\newblock
\urldef\tempurl%
\url{https://doi.org/10.1109/isdfs60797.2024.10527236}
\showDOI{\tempurl}


\bibitem[\protect\citeauthoryear{Liu, Wang, and Chen}{Liu
  et~al\mbox{.}}{2008}]%
        {liu2008multistep}
\bibfield{author}{\bibinfo{person}{Z. Liu}, \bibinfo{person}{C. Wang}, {and}
  \bibinfo{person}{S. Chen}.} \bibinfo{year}{2008}\natexlab{}.
\newblock \showarticletitle{Correlating Multi-Step Attack and Constructing
  Attack Scenarios Based on Attack Pattern Modeling}. In
  \bibinfo{booktitle}{\emph{Proceedings of the 2008 International Conference on
  Information Security and Assurance (ISA 2008)}}. \bibinfo{pages}{214--219}.
\newblock
\urldef\tempurl%
\url{https://doi.org/10.1109/ISA.2008.11}
\showDOI{\tempurl}


\bibitem[\protect\citeauthoryear{Micro}{Micro}{2014}]%
        {trendmicro2014sony}
\bibfield{author}{\bibinfo{person}{Trend Micro}.}
  \bibinfo{year}{2014}\natexlab{}.
\newblock \bibinfo{title}{The Hack of Sony Pictures: What We Know and What You
  Need to Know}.
\newblock \bibinfo{howpublished}{Trend Micro Security News}.
\newblock
\urldef\tempurl%
\url{https://www.trendmicro.com/vinfo/us/security/news/cyber-attacks/the-hack-of-sony-pictures-what-you-need-to-know}
\showURL{%
\tempurl}


\bibitem[\protect\citeauthoryear{Micro}{Micro}{2018}]%
        {trendmicro2018lazarus}
\bibfield{author}{\bibinfo{person}{Trend Micro}.}
  \bibinfo{year}{2018}\natexlab{}.
\newblock \bibinfo{title}{A Look into the Lazarus Group’s Operations}.
\newblock \bibinfo{howpublished}{Trend Micro Security News}.
\newblock
\urldef\tempurl%
\url{https://www.trendmicro.com/vinfo/us/security/news/cybercrime-and-digital-threats/a-look-into-the-lazarus-groups-operations}
\showURL{%
\tempurl}


\bibitem[\protect\citeauthoryear{Min, Ross, Sulem, Veyseh, Nguyen, Sainz,
  Agirre, Heintz, and Roth}{Min et~al\mbox{.}}{2023}]%
        {min2023nlp}
\bibfield{author}{\bibinfo{person}{B. Min}, \bibinfo{person}{H. Ross},
  \bibinfo{person}{E. Sulem}, \bibinfo{person}{A.~P. Veyseh},
  \bibinfo{person}{T.~H. Nguyen}, \bibinfo{person}{O. Sainz},
  \bibinfo{person}{E. Agirre}, \bibinfo{person}{I. Heintz}, {and}
  \bibinfo{person}{D. Roth}.} \bibinfo{year}{2023}\natexlab{}.
\newblock \showarticletitle{Recent Advances in Natural Language Processing via
  Large Pre-Trained Language Models: A Survey}.
\newblock \bibinfo{journal}{\emph{Comput. Surveys}} \bibinfo{volume}{5},
  \bibinfo{number}{2} (\bibinfo{date}{September} \bibinfo{year}{2023}).
\newblock
\urldef\tempurl%
\url{https://doi.org/10.1145/3605943}
\showDOI{\tempurl}


\bibitem[\protect\citeauthoryear{Mitra, Neupane, Chakraborty, Mittal, Piplai,
  Gaur, and Rahimi}{Mitra et~al\mbox{.}}{2024}]%
        {mitra2024localintel}
\bibfield{author}{\bibinfo{person}{S. Mitra}, \bibinfo{person}{S. Neupane},
  \bibinfo{person}{T. Chakraborty}, \bibinfo{person}{S. Mittal},
  \bibinfo{person}{A. Piplai}, \bibinfo{person}{M. Gaur}, {and}
  \bibinfo{person}{S. Rahimi}.} \bibinfo{year}{2024}\natexlab{}.
\newblock \showarticletitle{LOCALINTEL: Generating Organizational Threat
  Intelligence from Global and Local Cyber Knowledge}.
\newblock \bibinfo{journal}{\emph{arXiv}} (\bibinfo{date}{January}
  \bibinfo{year}{2024}).
\newblock
\urldef\tempurl%
\url{https://doi.org/10.48550/arxiv.2401.10036}
\showDOI{\tempurl}


\bibitem[\protect\citeauthoryear{Navarro, Deruyver, and Parrend}{Navarro
  et~al\mbox{.}}{2018}]%
        {navarro2018multistep}
\bibfield{author}{\bibinfo{person}{J. Navarro}, \bibinfo{person}{A. Deruyver},
  {and} \bibinfo{person}{P. Parrend}.} \bibinfo{year}{2018}\natexlab{}.
\newblock \showarticletitle{A Systematic Survey on Multi-Step Attack
  Detection}.
\newblock \bibinfo{journal}{\emph{Computers \& Security}}  \bibinfo{volume}{76}
  (\bibinfo{date}{July} \bibinfo{year}{2018}), \bibinfo{pages}{214--249}.
\newblock
\urldef\tempurl%
\url{https://doi.org/10.1016/j.cose.2018.03.001}
\showDOI{\tempurl}


\bibitem[\protect\citeauthoryear{Nonaka and Takeuchi}{Nonaka and
  Takeuchi}{1995}]%
        {nonaka1995knowledge}
\bibfield{author}{\bibinfo{person}{I. Nonaka} {and} \bibinfo{person}{H.
  Takeuchi}.} \bibinfo{year}{1995}\natexlab{}.
\newblock \bibinfo{booktitle}{\emph{The Knowledge-Creating Company: How
  Japanese Companies Create the Dynamics of Innovation}}.
\newblock \bibinfo{publisher}{Oxford University Press}, \bibinfo{address}{New
  York}.
\newblock
\urldef\tempurl%
\url{https://academic.oup.com/book/52097}
\showURL{%
\tempurl}


\bibitem[\protect\citeauthoryear{Polanyi}{Polanyi}{1966a}]%
        {polanyi1966logic}
\bibfield{author}{\bibinfo{person}{M. Polanyi}.}
  \bibinfo{year}{1966}\natexlab{a}.
\newblock \showarticletitle{The Logic of Tacit Inference}.
\newblock \bibinfo{journal}{\emph{Philosophy}} \bibinfo{volume}{41},
  \bibinfo{number}{155} (\bibinfo{year}{1966}), \bibinfo{pages}{1--18}.
\newblock
\urldef\tempurl%
\url{https://doi.org/10.1017/S0031819100066110}
\showDOI{\tempurl}


\bibitem[\protect\citeauthoryear{Polanyi}{Polanyi}{1966b}]%
        {polanyi1966tacit}
\bibfield{author}{\bibinfo{person}{M. Polanyi}.}
  \bibinfo{year}{1966}\natexlab{b}.
\newblock \bibinfo{booktitle}{\emph{The Tacit Dimension}}.
\newblock \bibinfo{publisher}{Doubleday \& Company, Inc.},
  \bibinfo{address}{New York}.
\newblock


\bibitem[\protect\citeauthoryear{Shashwat, Hahn, Ou, Goldgof, Hall, Ligatti,
  Rajagopalan, and Tabari}{Shashwat et~al\mbox{.}}{2024}]%
        {shashwat2024pentesting}
\bibfield{author}{\bibinfo{person}{K. Shashwat}, \bibinfo{person}{F. Hahn},
  \bibinfo{person}{X. Ou}, \bibinfo{person}{D. Goldgof}, \bibinfo{person}{L.
  Hall}, \bibinfo{person}{J. Ligatti}, \bibinfo{person}{S.~R. Rajagopalan},
  {and} \bibinfo{person}{A.~Z. Tabari}.} \bibinfo{year}{2024}\natexlab{}.
\newblock \showarticletitle{A Preliminary Study on Using Large Language Models
  in Software Pentesting}. In \bibinfo{booktitle}{\emph{Proceedings of the
  Workshop on SOC Operations and Construction (WOSOC 2024)}}.
  \bibinfo{address}{San Diego, CA, USA}, \bibinfo{pages}{1--7}.
\newblock
\urldef\tempurl%
\url{https://www.ndss-symposium.org/wp-content/uploads/wosoc2024-2-paper.pdf}
\showURL{%
\tempurl}


\bibitem[\protect\citeauthoryear{Shenoy and Mbaziira}{Shenoy and
  Mbaziira}{2024}]%
        {shenoy2024prompt}
\bibfield{author}{\bibinfo{person}{N. Shenoy} {and} \bibinfo{person}{A.~V.
  Mbaziira}.} \bibinfo{year}{2024}\natexlab{}.
\newblock \showarticletitle{An Extended Review: LLM Prompt Engineering in Cyber
  Defense}. In \bibinfo{booktitle}{\emph{Proceedings of the International
  Conference on Electrical, Computer and Energy Technologies}}.
  \bibinfo{address}{Sydney, Australia}, \bibinfo{pages}{1--6}.
\newblock
\urldef\tempurl%
\url{https://doi.org/10.1109/icecet61485.2024.10698605}
\showDOI{\tempurl}


\bibitem[\protect\citeauthoryear{Siracusano, Sanvito, Gonzalez, Srinivasan,
  Kamatchi, Takahashi, Kawakita, Kakumaru, and Bifulco}{Siracusano
  et~al\mbox{.}}{2023}]%
        {siracusano2023cti}
\bibfield{author}{\bibinfo{person}{G. Siracusano}, \bibinfo{person}{D.
  Sanvito}, \bibinfo{person}{R. Gonzalez}, \bibinfo{person}{M. Srinivasan},
  \bibinfo{person}{S. Kamatchi}, \bibinfo{person}{W. Takahashi},
  \bibinfo{person}{M. Kawakita}, \bibinfo{person}{T. Kakumaru}, {and}
  \bibinfo{person}{R. Bifulco}.} \bibinfo{year}{2023}\natexlab{}.
\newblock \showarticletitle{Time for aCTIon: Automated Analysis of Cyber Threat
  Intelligence in the Wild}.
\newblock \bibinfo{journal}{\emph{arXiv}} (\bibinfo{date}{July}
  \bibinfo{year}{2023}).
\newblock
\urldef\tempurl%
\url{https://doi.org/10.48550/arXiv.2307.10214}
\showDOI{\tempurl}


\bibitem[\protect\citeauthoryear{Soroush, Albanese, Mehrabadi, Iganibo, Mosko,
  Gao, Fritz, Rane, and Bier}{Soroush et~al\mbox{.}}{2020}]%
        {soroush2020sciborg}
\bibfield{author}{\bibinfo{person}{H. Soroush}, \bibinfo{person}{M. Albanese},
  \bibinfo{person}{M.~A. Mehrabadi}, \bibinfo{person}{I. Iganibo},
  \bibinfo{person}{M. Mosko}, \bibinfo{person}{J. Gao}, \bibinfo{person}{D.
  Fritz}, \bibinfo{person}{S. Rane}, {and} \bibinfo{person}{E. Bier}.}
  \bibinfo{year}{2020}\natexlab{}.
\newblock \showarticletitle{SCIBORG: Secure Configurations for the IoT Based on
  Optimization and Reasoning on Graphs}. In
  \bibinfo{booktitle}{\emph{Proceedings of the 8th IEEE Conference on
  Communications and Network Security (IEEE CNS 2020)}}. \bibinfo{pages}{10
  pages}.
\newblock
\urldef\tempurl%
\url{https://doi.org/10.1109/CNS48642.2020.9162256}
\showDOI{\tempurl}


\bibitem[\protect\citeauthoryear{Spradley}{Spradley}{2016}]%
        {spradley2016participant}
\bibfield{author}{\bibinfo{person}{J.~P. Spradley}.}
  \bibinfo{year}{2016}\natexlab{}.
\newblock \bibinfo{booktitle}{\emph{Participant Observation}}.
\newblock \bibinfo{publisher}{Waveland Press}.
\newblock
\urldef\tempurl%
\url{https://www.waveland.com/browse.php?t=689}
\showURL{%
\tempurl}


\bibitem[\protect\citeauthoryear{Sundaramurthy, Bardas, Case, Ou, Wesch,
  McHugh, and Rajagopalan}{Sundaramurthy et~al\mbox{.}}{2015}]%
        {sundaramurthy2015human}
\bibfield{author}{\bibinfo{person}{S.~C. Sundaramurthy}, \bibinfo{person}{A.~G.
  Bardas}, \bibinfo{person}{J. Case}, \bibinfo{person}{X. Ou},
  \bibinfo{person}{M. Wesch}, \bibinfo{person}{J. McHugh}, {and}
  \bibinfo{person}{S.~R. Rajagopalan}.} \bibinfo{year}{2015}\natexlab{}.
\newblock \showarticletitle{A Human Capital Model for Mitigating Security
  Analyst Burnout}. In \bibinfo{booktitle}{\emph{Proceedings of the 11th
  Symposium on Usable Privacy and Security (SOUPS)}}. \bibinfo{address}{Ottawa,
  Canada}, \bibinfo{pages}{347--359}.
\newblock
\urldef\tempurl%
\url{https://www.usenix.org/system/files/conference/soups2015/soups15-paper-sundaramurthy.pdf}
\showURL{%
\tempurl}


\bibitem[\protect\citeauthoryear{Sundaramurthy, McHugh, Ou, Rajagopalan, and
  Wesch}{Sundaramurthy et~al\mbox{.}}{2014}]%
        {sundaramurthy2014anthropological}
\bibfield{author}{\bibinfo{person}{S.~C. Sundaramurthy}, \bibinfo{person}{J.
  McHugh}, \bibinfo{person}{X. Ou}, \bibinfo{person}{S.~R. Rajagopalan}, {and}
  \bibinfo{person}{M. Wesch}.} \bibinfo{year}{2014}\natexlab{}.
\newblock \showarticletitle{An Anthropological Approach to Studying CSIRTs}.
\newblock \bibinfo{journal}{\emph{IEEE Security \& Privacy}}
  \bibinfo{volume}{12}, \bibinfo{number}{5} (\bibinfo{date}{Sept./Oct.}
  \bibinfo{year}{2014}), \bibinfo{pages}{41--47}.
\newblock
\urldef\tempurl%
\url{https://doi.org/10.1109/MSP.2014.84}
\showDOI{\tempurl}


\bibitem[\protect\citeauthoryear{Sundaramurthy, McHugh, Ou, Wesch, Bardas, and
  Rajagopalan}{Sundaramurthy et~al\mbox{.}}{2016}]%
        {sundaramurthy2016turning}
\bibfield{author}{\bibinfo{person}{S.~C. Sundaramurthy}, \bibinfo{person}{J.
  McHugh}, \bibinfo{person}{X. Ou}, \bibinfo{person}{M. Wesch},
  \bibinfo{person}{A.~G. Bardas}, {and} \bibinfo{person}{S.~R. Rajagopalan}.}
  \bibinfo{year}{2016}\natexlab{}.
\newblock \showarticletitle{Turning Contradictions into Innovations or: How We
  Learned to Stop Whining and Improve Security Operations}. In
  \bibinfo{booktitle}{\emph{Proceedings of the 12th Symposium on Usable Privacy
  and Security (SOUPS 2016)}}. \bibinfo{address}{Denver, CO, USA},
  \bibinfo{pages}{237--251}.
\newblock
\urldef\tempurl%
\url{https://www.usenix.org/system/files/conference/soups2016/soups2016-paper-sundaramurthy.pdf}
\showURL{%
\tempurl}


\bibitem[\protect\citeauthoryear{Thomas}{Thomas}{2006}]%
        {thomas2006inductive}
\bibfield{author}{\bibinfo{person}{D.~R. Thomas}.}
  \bibinfo{year}{2006}\natexlab{}.
\newblock \showarticletitle{A General Inductive Approach for Analyzing
  Qualitative Evaluation Data}.
\newblock \bibinfo{journal}{\emph{American Journal of Evaluation}}
  \bibinfo{volume}{27}, \bibinfo{number}{2} (\bibinfo{year}{2006}),
  \bibinfo{pages}{237--246}.
\newblock
\urldef\tempurl%
\url{https://doi.org/10.1177/1098214005283748}
\showDOI{\tempurl}


\bibitem[\protect\citeauthoryear{Touvron, Lavril, Izacard, Martinet, Lachaux,
  Lacroix, Rozière, Goyal, Hambro, and Azhar}{Touvron et~al\mbox{.}}{2023}]%
        {touvron2023llama}
\bibfield{author}{\bibinfo{person}{H. Touvron}, \bibinfo{person}{T. Lavril},
  \bibinfo{person}{G. Izacard}, \bibinfo{person}{X. Martinet},
  \bibinfo{person}{M.-A. Lachaux}, \bibinfo{person}{T. Lacroix},
  \bibinfo{person}{B. Rozière}, \bibinfo{person}{N. Goyal},
  \bibinfo{person}{E. Hambro}, {and} \bibinfo{person}{F. Azhar}.}
  \bibinfo{year}{2023}\natexlab{}.
\newblock \showarticletitle{LLaMA: Open and Efficient Foundation Language
  Models}.
\newblock \bibinfo{journal}{\emph{arXiv}} (\bibinfo{year}{2023}).
\newblock
\urldef\tempurl%
\url{https://doi.org/10.48550/arXiv.2302.13971}
\showDOI{\tempurl}


\bibitem[\protect\citeauthoryear{Trubowitz and Watanabe}{Trubowitz and
  Watanabe}{2021}]%
        {trubowitz2021geopolitical}
\bibfield{author}{\bibinfo{person}{P. Trubowitz} {and} \bibinfo{person}{K.
  Watanabe}.} \bibinfo{year}{2021}\natexlab{}.
\newblock \showarticletitle{The Geopolitical Threat Index: A Text-Based
  Computational Approach to Identifying Foreign Threats}.
\newblock \bibinfo{journal}{\emph{International Studies Quarterly}}
  \bibinfo{volume}{65}, \bibinfo{number}{3} (\bibinfo{date}{May}
  \bibinfo{year}{2021}), \bibinfo{pages}{852--865}.
\newblock
\urldef\tempurl%
\url{https://doi.org/10.1093/isq/sqab029}
\showDOI{\tempurl}


\bibitem[\protect\citeauthoryear{Wei, Wang, Schuurmans, Bosma, Ichter, Xia,
  Chi, Le, and Zhou}{Wei et~al\mbox{.}}{2022}]%
        {wei2022chain}
\bibfield{author}{\bibinfo{person}{J. Wei}, \bibinfo{person}{X. Wang},
  \bibinfo{person}{D. Schuurmans}, \bibinfo{person}{M. Bosma},
  \bibinfo{person}{B. Ichter}, \bibinfo{person}{F. Xia}, \bibinfo{person}{E.
  Chi}, \bibinfo{person}{Q.~V. Le}, {and} \bibinfo{person}{D. Zhou}.}
  \bibinfo{year}{2022}\natexlab{}.
\newblock \showarticletitle{Chain-of-Thought Prompting Elicits Reasoning in
  Large Language Models}. In \bibinfo{booktitle}{\emph{Advances in Neural
  Information Processing Systems}}, Vol.~\bibinfo{volume}{35}.
  \bibinfo{pages}{24824--24837}.
\newblock
\urldef\tempurl%
\url{https://doi.org/10.5555/3600270.3602070}
\showDOI{\tempurl}


\bibitem[\protect\citeauthoryear{Xiang and Wang}{Xiang and Wang}{2019}]%
        {xiang2019survey}
\bibfield{author}{\bibinfo{person}{W. Xiang} {and} \bibinfo{person}{B. Wang}.}
  \bibinfo{year}{2019}\natexlab{}.
\newblock \showarticletitle{A Survey of Event Extraction from Text}.
\newblock \bibinfo{journal}{\emph{IEEE Access}}  \bibinfo{volume}{7}
  (\bibinfo{year}{2019}), \bibinfo{pages}{173111--173137}.
\newblock
\urldef\tempurl%
\url{https://doi.org/10.1109/access.2019.2956831}
\showDOI{\tempurl}


\bibitem[\protect\citeauthoryear{Yang, Jin, Tang, and Tang}{Yang
  et~al\mbox{.}}{2024}]%
        {yang2024harnessing}
\bibfield{author}{\bibinfo{person}{J. Yang}, \bibinfo{person}{H. Jin},
  \bibinfo{person}{R. Tang}, {and} \bibinfo{person}{J. Tang}.}
  \bibinfo{year}{2024}\natexlab{}.
\newblock \showarticletitle{Harnessing the Power of LLMs in Practice: A Survey
  on ChatGPT and Beyond}.
\newblock \bibinfo{journal}{\emph{Commun. ACM}} \bibinfo{volume}{67},
  \bibinfo{number}{1} (\bibinfo{date}{January} \bibinfo{year}{2024}),
  \bibinfo{pages}{70--79}.
\newblock
\urldef\tempurl%
\url{https://doi.org/10.1145/3649506}
\showDOI{\tempurl}


\end{thebibliography}
